\documentclass[aps,prd,twocolumn,nofootinbib,superscriptaddress]{revtex4}
\usepackage[utf8]{inputenc}
\usepackage{graphicx}
\usepackage{amsmath,amssymb}
\usepackage{amsfonts}
\usepackage{physics}
\usepackage{xspace}
\usepackage{bm}
\usepackage{mathrsfs}
\usepackage{nicefrac}
\usepackage[dvipsnames]{xcolor}
\usepackage[unicode]{hyperref}
\hypersetup{colorlinks=true, citecolor=MidnightBlue, linkcolor=Blue, urlcolor=CornflowerBlue, linktocpage=true}
\usepackage[all]{hypcap}
\usepackage{multirow}
\usepackage[format=plain,justification=RaggedRight,labelsep=endash,font=small,labelfont=sc]{caption}
\usepackage[export]{adjustbox}

\definecolor {darkgreen}{rgb}{0.2,0.7,0.2}

\begin{document}
\title{Diffusion matrix associated with the diffusion processes of multiple conserved charges in a hot and dense hadronic matter}

\author{Arpan Das}
\email{arpan.das@ifj.edu.pl}
\affiliation{Institute  of  Nuclear  Physics  Polish  Academy  of  Sciences,  PL-31-342  Krak\'ow,  Poland}
\author{Hiranmaya Mishra}
\email{hm@prl.res.in}
\affiliation{Theory Division, Physical Research Laboratory, Navrangpura, Ahmedabad 380 009, India}
\author{Ranjita K. Mohapatra}
\email{ranjita.iop@gmail.com}
\affiliation{Department of Physics, Banki Autonomous College, Cuttack 754008, India}

\begin{abstract}
{Bulk matter produced in heavy ion collisions has multiple
 conserved quantum numbers like baryon number, strangeness and electric charge. The diffusion process
of these charges can be described by a diffusion matrix describing the interdependence of diffusion of
different charges. The diffusion coefficient matrix is estimated here from the Boltzmann kinetic theory
for the hadronic phase within relaxation time approximation. In the derivation for the same, we impose the Landau-Lifshitz
conditions of fit. This leads to e.g. the diagonal diffusion coefficients
to be manifestly positive definite. The explicit calculations are performed 
 within the ambit of hadron resonance gas model with and without excluded  volume corrections. It is seen that the off-diagonal
components can be significant to affect the charge diffusion in a fluid with 
 multiple conserved charges. The excluded volume correction effects is seen to be not significant in the estimation of the
elements of the diffusion matrix.}

\end{abstract}

\pacs{}
\date{\today \hspace{0.2truecm}}

\maketitle
\flushbottom

\section{Introduction}
\label{sec:intro}
Relativistic heavy-ion collision experiments, e.g. the Relativistic Heavy Ion Collider (RHIC) and 
the Large Hadron Collider (LHC) give us an unique opportunity to understand the properties of strongly 
interacting matter governed by the laws of  Quantum Chromodynamics (QCD). Experimental data at RHIC and LHC 
indicates a transient phase of deconfined quarks and gluons also known as the quark-gluon plasma (QGP). 
Quark-gluon plasma produced in the initial stage of high multiplicity heavy-ion collision undergoes 
subsequent space-time evolution and eventually results in a chemically and thermally equilibrated hadronic medium. 
The strongly interacting medium so produced shows collective motion and relativistic hydrodynamics along with 
a modeling of the early stage and freeze-out of hadrons have become the important tool for modeling 
relativistic heavy-ion collisions~\cite{Romatschke:2017ejr,Florkowski:2010zz,Gale:2013da,Jeon:2015dfa,Jaiswal:2016hex}.
 Various dissipative effects and the related transport coefficients, e.g., shear ($\eta$) and bulk ($\zeta$) viscosity, etc., can play a significant role in the hydrodynamical evolution of the strongly interacting medium.
Indeed, it has been argued that a small value of the kinematic viscosity i.e. small value of shear viscosity 
to entropy density ratio $(\eta/s)$, which is in accordance with the Kovtun-Son-Starinet (KSS) 
bound of $(\eta/s)=1/(4\pi)$ obtained using gauge gravity duality (AdS/CFT correspondence), can explain the flow data
  ~\cite{Heinz:2013th,Romatschke:2017ejr,Kovtun:2004de}. It may be emphasized that QCD is not conformal 
in nature and the deviation of the conformality is encoded in the bulk viscosity $\zeta$ of the strongly 
interacting medium~\cite{Gavin:1985ph,Hosoya:1983xm,Dobado:2012zf,Sasaki:2008fg,Sasaki:2008um,Karsch:2007jc,Finazzo:2014cna,
Wiranata:2009cz,Jeon:1995zm}. Similar to the shear viscosity, bulk viscosity also plays a very important role 
in the viscous hydrodynamic description of the QCD medium~\cite{Noronha-Hostler:2013gga,Ryu:2015vwa,Ryu:2017qzn,
Vujanovic:2019yih}. Other than the shear and bulk viscosities, electrical conductivity ($\sigma_{el}$) also becomes 
important in the magneto-hydrodynamic description of the QCD matter. This is due to the fact that in the noncentral heavy-ion collision, a large magnetic field of the order of $m_{\pi}^2$ is also expected to be produced. For phenomenological manifestation of such magnetic fields on the strongly interacting medium requires that a strong magnetic field survives for at least a few Fermi proper time in the plasma. The crucial parameter that determines the time evolution of magnetic field in the medium is the electrical conductivity~\cite{Tuchin:2010gx,Tuchin:2010vs,Inghirami:2016iru,Das:2017qfi,Greif:2016skc,Greif:2014oia,Puglisi:2014pda,Puglisi:2014sha,Cassing:2013iz,Steinert:2013fza,Aarts:2014nba,Aarts:2007wj,Amato:2013naa,Gupta:2003zh,Burnier:2012ts,Ding:2010ga,Kaczmarek:2013dya,Marty:2013ita}. Therefore these transport coefficients play a very important role for a comprehensive understanding of the hot and dense QCD medium produced in heavy-ion collisions.

The study of the QCD phase structure both at finite temperature and chemical potential is the goal of 
ultra relativistic heavy-ion collisions. While lattice QCD (LQCD) studies indicate the phase transition
from hadronic matter to quark gluon matter at high temperature and at vanishing or small
values of baryon chemical potential is a crossover~\cite{Aoki:2006we}, it is expected that at large baryon densities
 and low temperatures such a transition is most likely a first order transition
~\cite{Asakawa:1989bq,Ejiri:2008xt}. 
 Therefore, in the phase diagram of strong interaction in the plane of baryon chemical potential and temperature, one expects the first order line
ending in a critical end point (CEP). 
 Fluctuation of the conserved charges plays an important role to find the 
critical point~\cite{Stephanov:1999zu,Hatta:2003wn,Asakawa:2000wh,Jeon:1999gr,Ejiri:2005wq,Kitazawa:2013bta,Skokov:2012ds,Pal:2020ucy}. 
It has been argued that event-by-event fluctuations of conserved quantities such as net baryon number, net electric charge, and net strangeness are a possible signal of the QGP formation and quark-hadron phase transition~\cite{Asakawa:2000wh,Jeon:1999gr}. Moreover, due to the rapid expansion of the fireball fluctuations originated in the QGP phase may survive until the freeze-out and can be used as a signal of the QGP formation in the early stages of the relativistic heavy ion collisions~\cite{Asakawa:2000wh,Jeon:1999gr}. In the context of conserved charge fluctuations diffusion plays an important role as the time evolution of conserved charges is caused by diffusion process~\cite{Asakawa:2015ybt,Asakawa:2000wh,Jeon:2000wg,Shuryak:2000pd,Pratt:2019pnd}.

In nonrelativistic systems the diffusion process is described by Fick's law which relates the diffusion current 
$(\Delta\vec{J}_q)$  corresponding to  charge $q$, originated from the spatial inhomogeneity of the related
 charge density $n_q(t,\vec{x})$. Explicitly, $\Delta\vec{J}_q = -\kappa_q\vec{\nabla}n_q(t,\vec{x})$. 
The diffusion coefficient ($\kappa_q$) is a dissipative transport coefficient that characterizes the 
reaction strength of this thermal force~\cite{Fotakis:2019nbq}. If we consider baryon number fluctuation
 associated with the baryon number conservation then the effect of diffusion is not expected to be 
significant due to almost vanishing  net baryon density at mid rapidity region in ultra-relativistic heavy-ion 
collisions at RHIC and LHC~\cite{Monnai:2012jc}. 
However, with the Beam Energy Scan (BES) at RHIC, systematic low energy nuclear collisions have been 
going on to investigate the phase diagram of nuclear matter at finite net baryon density
~\cite{STAR:2010vob,Mohanty:2011nm,Mitchell:2012mx}. Furthermore, the heavy-ion collisions at 
Facility for Antiproton and Ion Research (FAIR) at Darmstadt and in Nuclotron-based Ion Collider 
fAcility (NICA) at Dubna one expects a baryon-rich medium~\cite{Odyniec:2013kna,STAR:2017sal,Friman:2011zz}. 
 With the decreasing collision energy, the net baryon density increases and the diffusion processes are
expected to play an increasingly important role in the dissipative dynamics of the evolution of hot and dense matter.

For a relativistic system the Fick's law can be expressed as, $\Delta J^i_q=\kappa _{qq}D^i(\mu_q/T)$.
Due to the presence of multiple conserved charges in QCD, i.e. baryon number ($B$), strangeness ($S$), 
electric charge ($Q$), etc., the simple Fick's law  as above, now gets modified. Since the strongly interacting 
particles, e.g. hadrons and quarks can carry more than one of these conserved charges, 
the diffusion current of each charge will no longer solely depends on the gradient of that specific charge. Since the 
gradients of every single charge density can generate a diffusion current of any other charge the diffusion currents of 
the conserved charges must get  coupled to each other.
Therefore, in the presence of multiple conserved charges, one has a generalized Fick's law,
\begin{align}
    \begin{pmatrix}
\Delta J^{i}_B \\
\Delta J^{i}_Q \\
\Delta J^{i}_S
\end{pmatrix}=\begin{pmatrix}
\kappa_{BB} & \kappa_{BQ} & \kappa_{BS} \\
\kappa_{QB} & \kappa_{QQ} & \kappa_{QS}\\
\kappa_{SB} & \kappa_{SQ} & \kappa_{SS}
\end{pmatrix}
    \begin{pmatrix}
D^{i}\alpha_B\\
D^{i}\alpha_Q \\
D^{i}\alpha_S
\end{pmatrix}.\nonumber
\end{align}
In the above, $\alpha_q=\mu_q/T$ with $q=B,S,Q$ and $D=u^\mu\partial_\mu$; $D^\mu=\partial ^\mu-u^\mu D$ and
$\Delta^{\mu\nu}=g^{\mu\nu}-u^\mu u^\nu$ is the projector orthogonal to fluid four velocity $u^\mu$. $\kappa_{qq^{\prime}}$ denotes the multicomponent diffusion matrix. 
The dynamics of the thermal potentials, i.e. $\alpha_B$, $\alpha_Q$, $\alpha_S$ and the corresponding currents in heavy-ion collisions are not often rigorously explored. 
It has been argued in Refs.~\cite{Monnai:2012jc,Greif:2017byw} that for low collision energies the influence of diffusion currents on the hydrodynamical evolution of the net-charge currents can be significant. Therefore not only the estimation of the diagonal components of the diffusion matrix elements, i.e. $\kappa_{BB}$, $\kappa_{QQ}$, $\kappa_{SS}$, but also the estimation of the off-diagonal terms, i.e.   $\kappa_{BQ}$, $\kappa_{QB}$, $\kappa_{BS}$, $\kappa_{SB}$, $\kappa_{QS}$, $\kappa_{SQ}$ are also very important.
In the present investigation we estimate the diffusion matrix element for the hadronic medium modeled by the 
hadron resonance gas model.

 At chemical freeze-out the hadron resonance gas (HRG) model describes quite well 
the hadronic phase of the strongly interacting medium created in heavy ion 
collisions~\cite{Braun-Munzinger:2003pwq,Andronic:2005yp}. In its simplest form assuming the freezeout
 of strange and nonstrange particles on same footing, HRG model has only two parameters $T$ and $\mu_B$. 
In general the thermodynamics of interacting hadrons can be nontrivial, however, in the
 presence of narrow resonances it has been argued that the thermodynamics of interacting gas of hadrons can
 be approximated by the non-interacting gas of hadrons and resonances~\cite{PhysRev.187.345,PhysRevD.10.694}. Due to
the small number of parameters and simple structure HRG model and its various extensions have been
 well explored regarding the experimental  result  of  the  thermal abundance of different particle
 ratios in the heavy ion collisions
~\cite{Braun-Munzinger:2001hwo,Cleymans:1999st,Becattini:2000jw,Cleymans:2004pp,Andronic:2008gu},
 thermodynamics~\cite{Karsch:2003zq,Braun-Munzinger:2015hba}, conserved charge fluctuations
~\cite{Nahrgang:2014fza,Bhattacharyya:2013oya,Garg:2013ata,PhysRevD.86.034509,Begun:2006jf},
 as well as transport coefficients for hadronic matter
~\cite{Greif:2016skc,Puglisi:2014pda,Puglisi:2014sha,Prakash:1993bt,Wiranata:2012br,
PhysRevC.83.014906,Khvorostukhin:2010aj,PhysRevC.86.054902,PhysRevC.77.024911,PhysRevC.86.024913,
PhysRevC.85.014908,PhysRevC.88.068201,PhysRevC.89.045201,Wiranata:2014kva,Wiranata:2012vv,
Tawfik:2010mb,PhysRevLett.103.172302,Kadam:2014cua,Kadam:2014xka,Ghosh:2014yea,Demir:2014kda,PhysRevC.90.025202,
PhysRevC.97.055204,PhysRevC.84.054911,PhysRevD.90.094014,Bass:1998ca,PhysRevC.92.035203,PhysRevD.99.014015,
Das:2021qii,Mohapatra:2019mcl}. 

Initial investigations on the diffusion matrix of the strongly interacting matter have been discussed 
in Refs.~\cite{Greif:2017byw,Fotakis:2021diq}. Explicit expressions of the diffusion matrix have been derived
 within the classical kinetic theory approach, both in the first order Chapman-Enskog expansion as well as in
 the relaxation time approximation. In the present investigation, we also follow the classical kinetic theory
 approach to obtain the expression of the diffusion matrix ($\kappa_{qq^{\prime}}$), keeping in mind 
the Landau-Lifshitz matching condition in the local rest frame. The expression of the diffusion matrix so obtained 
in the present work is significantly different from the expression of $\kappa_{qq^{\prime}}$ as given
 in Refs.~\cite{Greif:2017byw,Fotakis:2021diq}, e.g. the diagonal components of the diffusion matrix are
 not manifestly positive definite in Refs.~~\cite{Greif:2017byw,Fotakis:2021diq,Fotakis:2019nbq}.
 In the present work, we explicitly show that the diagonal component of the diffusion matrix ($\kappa_{qq}$) 
is positive definite. It should be noted that the positivity of various components is not so obvious.
 Signs of various transport coefficients in the context of dissipative hydrodynamics can be obtained by demanding
 entropy production. 

 The paper is organized in the following manner, in Sec.~\eqref{formalism} we discuss the formalism of diffusion matrix 
element within the framework of the Boltzmann kinetic equation and attempt to solve for the distribution functions within
 relaxation time approximation. In Sec.~\eqref{hrgmodels} we briefly discuss 
the hadron resonance gas model including the ideal HRG model as well as its extension to the excluded volume HRG model. 
In Sec.~\eqref{results} we present the estimation of diffusion coefficient for the hadron resonance gas model. Finally, 
we summarize and draw the conclusion of our investigations in Sec.~\eqref{conclusion}.

\section{Formalism}
\label{formalism}
Let us consider the covariant Boltzmann equation in the absence of external force, 
\begin{align}
& p_a\cdot \partial f_a(x,p) = \mathcal{C}_a = (u\cdot p_a)\sum_{b,c,d}\int \frac{d^3p_b}{(2\pi)^3} \frac{d^3p_c}{(2\pi)^3}\frac{d^3p_c}{(2\pi)^3}\nonumber\\
&~~~~~~~~~~~~~~~~~\times\frac{1}{1+\delta_{cd}}\left(f_cf_d-f_af_b\right)W(a,b|c,d),
\label{equ1ver1}
\end{align}
where, 
\begin{align}
    W(a,b|c,d) = \frac{(2\pi)^4\delta^4(p_a+p_b-p_c-p_d)}{2E_a 2E_b2E_c 2E_d}|\mathcal{M}(a,b|c,d)|^2.
    \label{equ2ver1}
\end{align}
Here $f_a$ is the out of equilibrium distribution function of species $`a'$, $\mathcal{C}$ is the collision term and $\mathcal{M}$ is the transition matrix element for the process $a+b\rightarrow c+d$. The factor $1/(1+\delta_{cd})$ takes into account the possibility of identical particles. 
The scalar product is the usual dot product $a\cdot b=a^\mu b_\mu=g_{\mu\nu}a^\mu b^\nu$; $g_{\mu\nu}$
is the flat space-time metric with signature $(+,-,-,-)$. Note that without introducing too many notations for momentum vector, throughout the manuscript we use the same notation for momentum four vector and the magnitude of the momentum three vector, i.e. $p^{\mu}\equiv(E_p,\vec{p})$ is the momentum four vector with $E_p=\sqrt{p^2+m^2}$ where $p\equiv|\vec{p}|$ is the magnitude of the momentum three vector.  

We consider the system under consideration to be initially in global equilibrium. Small perturbation of the single-particle distribution function from equilibrium is generated due to small gradients in the thermal potentials. This perturbation away from the equilibrium generates a diffusion current in the corresponding charges. 
To obtain a solution of the Boltzmann equation, we write the distribution function in powers
 of the Knudsen number and truncate 
such an expansion in the lowest order  as~\cite{Fotakis:2019nbq,Fotakis:2021diq,Chakraborty:2010fr}, 
\begin{align}
f_a(x,p)=f_a^{(0)}(p)\left(1+\phi^a(x,p)\right)
\label{equ3ver1}
\end{align}
where we have pushed the space-time dependence to the fluctuating part through  $\phi ^a$, and 
the equilibrium distribution function is given by,
\begin{align}
f_a^{(0)}& = g_a \exp(-\beta u\cdot p_a+\beta\sum_q q_a\mu_q)\nonumber\\
& = g_a\exp(-\beta u\cdot p_a+\alpha_a),
\label{equ4ver1}
\end{align}
where $u^\mu$ is the fluid four velocity with respect to which the invariant $p\cdot u$ is measured, $\beta=1/T$ 
is the inverse of temperature of the medium in the rest frame and $g_a$ is the degeneracy factor. Here, $\alpha^a=\sum_qq_a\alpha_q\equiv\beta\sum_q q_a\mu_q$, $\alpha_q=\beta\mu_q$ and $\mu_q$ corresponding to different chemical potentials (e.g. $q=B,S,Q$).

Next, to identify the transport coefficients, we write down the structure of the energy-momentum tensor $T^{\mu\nu}$
and the conserved current $J_q^\mu$ corresponding to a conserved charge $q$. In terms of temperature $T$,
chemical potential $\mu_q$ and flow velocity $u^\mu$, these are given as
\begin{align}
T^{\mu\nu}=-Pg^{\mu\nu}+\omega u^\mu u^\nu+\Delta T^{\mu\nu},
\end{align}
and
\begin{align}
J_q^{\mu}=n_qu^\mu+\Delta J_q^\mu.
\end{align}
We shall be taking $u^\mu$ as the velocity of energy flow normalized as $u_\mu u^\mu=1$. Further, in the above $P$, $\varepsilon$ is the pressure and energy density respectively
and  $\omega=\varepsilon+P$ is the enthalpy. $\Delta T^{\mu\nu}$ is the dissipative correction to the energy-momentum tensor due to viscosity,
\begin{align}
\Delta T^{\mu\nu}=\eta\left(D^\mu u^\nu+D^\nu u^\mu+\frac{2}{3} \Delta^{\mu\nu}\theta\right)-\zeta\Delta^{\mu\nu}\theta
\label{equ7ver1}
\end{align}
and, the dissipative contribution to the conserved current is given as,
\begin{align}
\Delta J_q^\mu=\kappa_{qq'}D^\mu \alpha_{q'},
\label{equ8ver1}
\end{align}
which is a relativistic form of Fick's law  generalised to different conserved charges $q$ with $\kappa_{qq'}$ being the diffusion matrix coefficient. The diffusion current is generated by the gradient in the thermal potential $\alpha_q$.
In the above, $\theta=\partial\cdot u$ is the expansion scalar; $D=u^\mu\partial_\mu$; $D^\mu=\partial ^\mu-u^\mu D$ and
$\Delta^{\mu\nu}=g^{\mu\nu}-u^\mu u^\nu$ is the projector orthogonal to $u^\mu$.

Having defined the transport coefficients from the structure of the energy-momentum tensor and the currents 
associated with the conserved charges, we next express these quantities in terms of the microscopic distribution 
functions obtained by solving the Boltzmann equation. The idea is to then identify the dissipation coefficients in 
terms of the distribution functions by comparing the corresponding tensor structures associated with 
the $T^{\mu\nu}$ and $J^\mu_q$.
The energy-momentum tensor is written as,
\begin{align}
T^{\mu\nu}=\sum_a\int \frac{d^3p_a}{(2\pi)^3}\frac{p^\mu_a p^\nu_a}{E_a}f_a,
\label{equ9ver1}
\end{align}
and the current corresponding to a conserved charge $q$ given as
\begin{align}
J_q^\mu=\sum_a q_a \int \frac{d^3p_a}{(2\pi)^3} \frac{p^{\mu}_a}{E_a}f_a. 
\label{equ10ver1}
\end{align}
Before proceeding further, we make a comment here, that in the present case, we are not considering here, the possibility
of including mean-field effects or the medium dependent masses which can be generalized easily using the methods
used in e.g. Refs.~\cite{Albright:2015fpa,Albright:2015edp,Deb:2016myz,Chakraborty:2010fr}. 

Now, let us note that the non-equilibrium part $\phi_a$ of the distribution function $f_a$ leads to the non-equilibrium
contributions $\Delta T^{\mu\nu}$ and $\Delta J^\mu_q$. This means $\phi_a$ should have the same tensor structure as $\Delta T^{\mu\nu}$ and $\Delta J^\mu_q$. Thus we take the form of $\phi_a$ as~\cite{Albright:2015fpa},
\begin{align}
\phi_a=-A_a\theta-\sum_q B_a^q p_a^\mu D_\mu\alpha_q+C_a p_a^\mu p_a^\nu \Sigma_{\mu\nu}
\label{phia}
\end{align}
where, the functions $A_a$, $B_a^q$ and $C_a$ are functions of magnitude of  momentum and $\Sigma_{\mu\nu}=
D_{\mu} u_\nu+D_\nu u_\mu+\frac{2}{3}\Delta_{\mu\nu}\theta$. This leads to, 
\begin{align}
    \Delta J^{\mu}_q = & \sum_a q_a\int \frac{d^3p_a}{(2\pi)^3}\frac{p_a^{\mu}}{E_a}f_a^{(0)}\phi_a,\nonumber\\
    = & \sum_a q_a\int \frac{d^3p_a}{(2\pi)^3}\frac{p_a^{\mu}}{E_a}f_a^{(0)}\nonumber\\
    & \times \bigg(-A_a\theta-\sum_{q^{\prime}} B_a^{q^{\prime}} p_a^\sigma D_\sigma\alpha_{q^{\prime}}+C_a p_a^\sigma p_a^\rho \Sigma_{\sigma\rho}\bigg).
    \label{equ12ver3}
\end{align}
In $\Delta J_q^i$, terms with $A_a$ and $C_a$ would give rise to odd momentum integration. Therefore, 

\begin{align}
    \Delta J_q^i = & \sum_a q_a \int \frac{d^3p_a}{(2\pi)^3}\frac{p_a^i}{E_a}f_a^{(0)}(-)B^{q^{\prime}}_ap_a^{\rho}D_{\rho}\alpha_{q^{\prime}}\nonumber\\
    = & \sum_a q_a \int \frac{d^3p_a}{(2\pi)^3}\frac{p_a^2}{3E_a}f_a^{(0)}B^{q^{\prime}}_a D^{i}\alpha_{q^{\prime}}\nonumber\\
    = & \kappa_{qq^{\prime}}D^{i}\alpha_{q^{\prime}}, 
\end{align}
here the diffusion matrix $\kappa_{qq^{\prime}}$ can be identified as, 
\begin{align}
    \kappa_{qq^{\prime}} =  \sum_a q_a \int \frac{d^3p_a}{(2\pi)^3}\frac{p_a^2}{3E_a}f_a^{(0)}B^{q^{\prime}
    }_a.
\label{equ14ver1}
\end{align}

To solve for the functions $A_a$, $B_a^q$, and $C_a$, we use the Boltzmann equation within the Chapman-Enskog approximation.
This corresponds to expanding both the sides of the Boltzmann equation to first order in $\phi_a$.
Let us start with the L.H.S of the Boltzmann Eq.~\eqref{equ1ver1}. Using $\partial_\mu=u_\mu D+D_\mu$, we have
\begin{align}
p_a^\mu\partial_\mu f_a^{(0)}= & p_a^{\mu}(u_{\mu}D+D_{\mu})f_a^{(0)}\nonumber\\
= & (E_a D+p^{\mu}_a D_{\mu})f_a^{(0)}.
\label{equ15ver1}
\end{align}
Using the expression of the equilibrium distribution function as given in Eq.~\eqref{equ4ver1} we obtain, 
\begin{align}
    Df_a^{(0)} = f_a^{(0)}\bigg(-E_aD\beta-\beta p_a^{\mu}D u_{\mu}+D\alpha^a\bigg).
    \label{equ16ver1}
\end{align}
Furthermore, $D u_\mu$ can be estimated by considering the conservation of energy-momentum tensor in ideal hydrodynamics, i.e. $\partial_\mu T_{(0)}^{\mu\nu}=0$ where,
\begin{align}
T^{\mu\nu}_{(0)}=(\varepsilon+P) u^\mu u^\nu-P g^{\mu\nu}.
\label{equ17ver1}
\end{align}
Taking the projection of $\partial_{\mu}T^{\mu\nu}_{(0)}=0$ in the direction orthogonal to $u^{\mu}$ one gets, 
\begin{align}
 & Du^{\alpha}=\frac{1}{\omega}D^{\alpha}P.
\label{equ18ver1}
\end{align}
Further, using the thermodynamics we obtain~\cite{Ollitrault:2007du}, 
\begin{align}
    D_{\alpha}\varepsilon = T D_{\alpha}s+\sum_q \mu_q D_{\alpha}n_q,
    \label{equ19ver1}
\end{align}
and, 
\begin{align}
    & P = -\varepsilon+Ts+\sum_q \mu_q n_q, \nonumber\\
\implies &  D_{\alpha}P=-D_{\alpha}\varepsilon+TD_{\alpha}s+(D_{\alpha}T)s \nonumber\\
& ~~~~~~~+ \sum_q \mu_q D_{\alpha}n_q+ \sum_q(D_{\alpha}\mu_q)n_q.
\label{equ20ver1}
\end{align}
Using Eqs.~\eqref{equ19ver1} and \eqref{equ20ver1} we get, 
\begin{align}
D_{\alpha}P    & = \frac{\omega}{T}D_{\alpha}T-\sum_q n_q\mu_q \frac{D_{\alpha}T}{T}+\sum_q n_q D_{\alpha}\mu_q.
\label{equ21ver1}
\end{align}
Note that, 
\begin{align}
& T\sum_q n_qD_{\alpha}\alpha_q = T\sum_q n_qD_{\alpha}(\mu_q/T)\nonumber\\
&~~~~~~~~~~~~ = \sum_q \left(n_q D_{\alpha}\mu_q-n_q\frac{\mu_q}{T}D_{\alpha}T\right).
\label{equ22ver1}
\end{align}
Therefore, Eqs.~\eqref{equ21ver1} and \eqref{equ22ver1} gives us, 
\begin{align}
    D_{\alpha}P=\frac{\omega}{T}D_{\alpha}T+\sum_q Tn_q D_{\alpha}\alpha_q,
\label{equ23ver1}
\end{align}
leading to,
\begin{align}
 & \beta Du^{\alpha}=-D^{\alpha}\beta + \sum_q\frac{n_q}{\omega}D^{\alpha}\alpha_q.
\label{equ24ver1}
\end{align}
Using the expression of $Du^{\alpha}$ as given in Eq.~\eqref{equ24ver1} in Eq.~\eqref{equ16ver1} we get, 
\begin{align}
    Df_a^{(0)} & =-f_a^{(0)}\bigg(E_a D\beta+p_a^{\mu}\bigg(-D_{\mu}\beta\nonumber\\
    & ~~~~~~~~~~~~~~+\sum_q\frac{n_q}{\omega}D_{\mu}\alpha_q\bigg)-D\alpha^a\bigg).
\label{equ25ver1}
\end{align}
Again, 
\begin{align}
    & D_{\mu}f_a^{(0)}=f_a^{(0)}\bigg(-D_{\mu}\beta(u\cdot p_a)-\beta p^{\alpha}_aD_{\mu}u_{\alpha}+D_{\mu}\alpha^a\bigg).
\label{equ26ver1}
\end{align}
Therefore, 
\begin{align}
    p_a^{\mu}D_{\mu}f_a^{(0)} 
    & = -f_a^{(0)}\bigg(E_a p_a^{\mu}D_{\mu}\beta+\beta p^{\mu}_ap^{\alpha}_aD_{\mu}u_{\alpha}\nonumber\\
    &~~~~~~~~~~~~~~~~~-p^{\mu}_aD_{\mu}\alpha^a \bigg).
\label{equ27ver1}
\end{align}

The second term within the parenthesis on the right hand side can be expressed as, 
\begin{align}
    & p_a^{\mu}p_a^{\alpha}D_{\mu}u_{\alpha}\nonumber\\
    = & \frac{1}{2}p_a^{\mu}p_a^{\alpha}\bigg(D_{\mu}u_{\alpha}+D_{\alpha}u_{\mu}-\frac{2}{3}\Delta_{\alpha\mu}\theta\bigg)+\frac{1}{3}p_a^{\mu}p_a^{\alpha}\Delta_{\alpha\mu}\theta\nonumber\\
    = & \frac{1}{2}p_a^{\mu}p_a^{\alpha} \Sigma_{\mu\alpha}+\frac{1}{3}p_a^{\mu}p_a^{\alpha}\Delta_{\alpha\mu}\theta,
    \label{equ28ver1}
\end{align}
so that, 
\begin{align}
    p_a^{\mu}D_{\mu}f_a^{(0)}
    & = -f_a^{(0)}\bigg(E_a p_a^{\mu}D_{\mu}\beta+\frac{\beta}{2}p_a^{\mu}p_a^{\alpha} \Sigma_{\mu\alpha}\nonumber\\
    &+\frac{\beta}{3}p_a^{\mu}p_a^{\alpha}\Delta_{\alpha\mu}\theta-p^{\mu}_aD_{\mu}\alpha^a \bigg),
\label{equ29ver1}
\end{align}
using Eqs.~\eqref{equ25ver1} and \eqref{equ29ver1}, Eq.~\eqref{equ15ver1} can be simplified to, 
\begin{align}
 p_a^{\mu}\partial_{\mu}f_a^{(0)}  & =E_aDf_a^{(0)}+p_a^{\mu}D_{\mu}f_{a}^{(0)}\nonumber\\
&= -f_a^{(0)}\bigg[E_a^2D\beta-E_aD\alpha^a\nonumber\\
&~~ +\beta p_a^{\mu}p_{a}^{\alpha}\bigg(\frac{1}{2}\Sigma_{\mu\alpha}+\frac{1}{3}\Delta_{\mu\alpha}\theta\bigg)\nonumber\\
&~~ + p_a^{\mu}\sum_q \bigg(\frac{E_a n_q}{\omega}-q_a\bigg)D_{\mu}\alpha_q\bigg].
\label{equ30ver1}
\end{align}


In the following, we shall confine our attention to diffusion only. In such a case, only the third term in the square bracket of Eq.~\eqref{equ30ver1} will be relevant and we write,
\begin{align}
\frac{p_a^{\mu}\partial_{\mu}f_a^{(0)}}{E_a}\simeq-f_a^{(0)}\frac{p_a^{\mu}}{E_a}\sum_q \bigg(\frac{E_a n_q}{\omega}-q_a\bigg)D_{\mu}\alpha_q.
\label{equ31ver1}
\end{align}
Next let us consider the R.H.S of the Boltzmann equation (Eq.~\eqref{equ1ver1}). In the Chapman-Enskog approximation, the collision term can be written as,
\begin{align}
\mathcal{C}_a= & E_af_a^{(0)}\sum_{b,c,d}\frac{1}{1+\delta_{cd}}\int \frac{d^3p_b}{(2\pi)^3} \frac{d^3p_c}{(2\pi)^3} \frac{d^3p_d}{(2\pi)^3}f_b^{(0)}\nonumber\\
& W(a,b|c,d)\left(\phi_c+\phi_d-\phi_a-\phi_b\right).
\label{equ32ver1}
\end{align}
Let us note that the collision term vanishes in equilibrium due to the detailed balance condition which follows from the energy momentum conserving delta function. 
Therefore in equilibrium $f_a^{(0)}f_b^{(0)}=f_c^{(0)}f_d^{(0)}$. As for the L.H.S, we shall confine our attention to diffusive process only so that $\phi_a\simeq-\sum_q B_a^q p_a^\mu D_\mu\alpha_q$.
Equating the coefficients of
$D_{\mu}\alpha_q$ from Eqs.~\eqref{equ31ver1} and \eqref{equ32ver1} leads to,
\begin{align}
& \frac{p_a^\mu}{E_a}\left(\frac{E_qn_q}{\omega}-q_a\right)= \sum_{b,c,d}\int \frac{d^3p_b}{(2\pi)^3} \frac{d^3p_c}{(2\pi)^3} \frac{d^3p_d}{(2\pi)^3} f_b^{(0)}\nonumber\\
& ~~~~~~~\frac{1}{1+\delta_{cd}}W(a,b|c,d)(B_c^qp^{\mu}_c+B_d^qp^{\mu}_d-B_a^qp^{\mu}_a-B_b^qp^{\mu}_b).
\label{equ33ver1}
\end{align}
The above is an integral equation for the functions $B_a^q$'s which depend upon the magnitude of the momentum $p_a$.

Let us note that if we have a particular solution, denoted as $B_{a-part}^q$, we can generate another 
solution $B_{a-part}^q-b^q $,
where $b_q$ is a constant independent of the species $a$. The reason is that the scattering term conserves the charge number.
 This can be easily seen by making a substitution $B_a^q=B_{a-part}^q-b^q $ 
in Eq.\eqref{equ33ver1}. 
This calls for additional constraints to select an unique solution.
This arbitrariness is associated with the freedom to define the local rest frame or, 
equivalently, the flow velocity $u^{\mu}$. In the Landau Lifshitz definition, $u^{\mu}$ is the velocity of the flow of energy.
On the other hand,  the Eckart definition of flow velocity corresponds to flow of conserved charge.
Choosing a specific frame is called the condition of fit. To choose the Landau Lifshitz frame, we impose 
the condition in the local rest frame as~\cite{Albright:2015edp,Albright:2015fpa}
\begin{align}
\Delta T^{0i}=0,\quad\quad \Delta J^0_q=0.
\end{align}

The second condition is trivially satisfied in the local rest frame, which can be understood from Eq.~\eqref{equ12ver3}. 
In the local rest frame $u^{\mu}=(1,0,0,0)$, so that $\theta$ and $\Sigma^{\mu\nu}$ vanish in the local rest frame. 
Furthermore, in the local rest frame $D^{\mu}$ operator has only spatial components. This
 makes the integrand of $\Delta J^{0}_q$ an odd function of momentum. Therefore $\Delta J^0_q=0$ in the local rest frame.

The first condition, i.e. $\Delta T^{0i}=0$  means
\begin{align}
\sum_q\sum_a\int \frac{d^3p_a}{(2\pi)^3} p_a^i (-)B_a^q p_a^j D_j\alpha_q f_a^{(0)}=0,
\end{align}
With $B_a^q=B_{a-part}^q-b^q$, the above condition reduces to,
\begin{align}
& \sum_q\sum_a\int \frac{d^3p_a}{(2\pi)^3} p_a^i(B_{a-part}^q-b^q) p_a^j D_j\alpha_q f_a^{(0)}=0,\nonumber\\
 \implies & \sum_q b_q \sum_a\int \frac{d^3p_a}{(2\pi)^3} p_a^i p_a^j D_j\alpha_q f_a^{(0)} \nonumber\\
 & = \sum_q\sum_a\int \frac{d^3p_a}{(2\pi)^3} p_a^i p_a^j B_{a-part}^q D_j\alpha_q f_a^{(0)}
\end{align}
Comparing the coefficient of $D_j\alpha_q$ from both sides of the above equation we obtain, 
\begin{align}
    b_q\sum_a \int \frac{d^3p_a}{(2\pi)^3} p_a^2 f_a^{(0)}=\sum_a\int \frac{d^3p_a}{(2\pi)^3}p_a^2 B_{a-part}^q f_a^{(0)}.
\end{align}
The quantity multiplying $b_q$ in the L.H.S of the above equation is just $3T \omega$ (for a detailed derivation see Appendix~\eqref{appenA}). This leads to,
\begin{align}
    b^q=\frac{1}{3T\omega}\sum_a\int \frac{d^3p_a}{(2\pi)^3}p_a^2 B_{a-part}^q f_a^{(0)}.
\label{equ38ver1}
\end{align}
Now in Eq.\eqref{equ14ver1}, we substitute $B_a^q=B_{a-part}^q-b^q$, so that the diffusion coefficient is given by,
\begin{align}
    \kappa_{qq^{\prime}} = & \sum_a q_a \int \frac{d^3p_a}{(2\pi)^3}\frac{p_a^2}{3E_a}f_a^{(0)}\left(B_{a-part}^{q^\prime}-b^{q^\prime}\right)\nonumber\\
    = & \sum_a q_a \int \frac{d^3p_a}{(2\pi)^3}\frac{p_a^2}{3E_a}f_a^{(0)} B_{a-part}^{q^\prime}\nonumber\\
    & - b^{q^\prime}\sum_a q_a \int \frac{d^3p_a}{(2\pi)^3}\frac{p_a^2}{3E_a}f_a^{(0)}.
\label{equ39ver1}
\end{align}
Further, it can be shown that the factor multiplying $b^{q^\prime}$ in Eq.~\eqref{equ39ver1} is $n_q T$ (for a detailed derivation see Appendix~\eqref{appenB}), i.e. \begin{align}
    \sum_a q_a \int \frac{d^3p_a}{(2\pi)^3}\frac{p_a^2}{3E_a}f_a^{(0)} = n_q T. 
    \label{equ40ver1}
\end{align}
Using Eqs.~\eqref{equ38ver1} and \eqref{equ40ver1} in Eq.~\eqref{equ39ver1} we obtain, 
\begin{align}
    & \kappa_{qq^{\prime}}  =  \sum_a q_a \int \frac{d^3p_a}{(2\pi)^3}\frac{p_a^2}{3E_a}f_a^{(0)} B_{a-part}^{q^\prime}\nonumber\\
    & ~~~~~~~~-\frac{n_q}{3\omega} \sum_a q_a \int \frac{d^3p_a}{(2\pi)^3} p_a^2 f_a^{(0)} B_{a-part}^{q^\prime}\nonumber\\
    & = \sum_a\int \frac{d^3p_a}{(2\pi)^3}\frac{p_a^2}{3E_a}\left(q_a-\frac{n_qE_a}{\omega}\right) f_a^{(0)} B_{a-part}^{q^\prime}.
    \label{equ41ver1}
\end{align}
Thus once we know $B_{a-part}^{q'}$, we have the expression for the
diffusivity as above. In this study, we use the relaxation time approximation, where it is assumed that all particles are in equilibrium except for the species $`a'$ in the Boltzmann equation for $f_a$, i.e $B_{b-part}=0=B_{c-part}=B_{d-part}$ in Eq.~\eqref{equ33ver1} then
we have,
\begin{align}
    \bigg(q_a-\frac{E_an_q}{\omega}\bigg)\frac{p_a^{\mu}}{E_a} & =\frac{1}{1+\delta_{cd}}\int \frac{d^3p_b}{(2\pi)^3} \frac{d^3p_c}{(2\pi)^3} \frac{d^3p_d}{(2\pi)^3}\nonumber\\
    & ~~~\times f_b^{(0)}W(a,b|c,d)B_{a-part}^qp_a^{\mu}.
    \label{equ42ver1}
\end{align}
In the above equation, we can identify the energy-dependent relaxation time of particle $`a'$ as, 
\begin{align}
    \tau_a^{-1}(E_a)=\frac{1}{1+\delta_{cd}}\int \frac{d^3p_b}{(2\pi)^3} \frac{d^3p_c}{(2\pi)^3} \frac{d^3p_d}{(2\pi)^3}f_b^{(0)}W(a,b|c,d).
\end{align}
Therefore from Eq.~\eqref{equ42ver1} we get, 
\begin{align}
    B_{a-part}^q=\frac{\tau_a}{E_a}\bigg(q_a-\frac{E_an_q}{\omega}\bigg).
\end{align}

Substituting the above in the expression for $\kappa_{qq^{\prime}}$ as in Eq.\eqref{equ41ver1} we have,
\begin{align}
    & \kappa_{qq^{\prime}}
     = \sum_a\int \frac{d^3p_a}{(2\pi)^3}\frac{p_a^2}{3E_a^2}\left(q_a-\frac{n_qE_a}{\omega}\right) \nonumber\\
     &~~~~~~~~~~~~~~~~~~~~ \times \tau_a\left(q^{\prime}_a-\frac{n_q^{\prime}E_a}{\omega}\right) f_a^{(0)}.
     \label{equ45ver1}
\end{align}
This makes the expression of $\kappa_{qq}$ positive definite. Also note that $\kappa_{qq^{\prime}}$ is symmetric with respect to the change $q\leftrightarrow q^{\prime}$~\cite{PhysRev.37.405,PhysRev.38.2265,Fotakis:2019nbq}. A few comments about the various approximation methods in the calculation of various transport coefficients are in order here. Several methods, e.g. the Relaxation Time Approximation (RTA), the Chapman-Enskog (CE) method, etc. have been discussed in the literature for the calculation of the transport coefficients.
The Chapman-Enskog approach is a variational approach and in this method depending on the order of approximation used one can obtain solutions with an arbitrary accuracy~\cite{Wiranata:2012br}. The relaxation time approach on the other hand is based on an ansatz for the collision kernel in the Boltzmann equation. Such an ansatz for the collision kernel does not allow to have control
of the precision of the approximation. The relaxation time approximation is incompatible with microscopic and macroscopic conservation laws. Novel approaches have been proposed to overcome such problems, e.g.~\cite{PhysRev.94.511,Rocha:2021zcw}. Nonetheless, the qualitative nature of various transport coefficients as obtain in the relaxation time approximation is similar to the Chapman-Enskog approach and the RTA approach has been even more often used due to its simplicity. Various quantities, e.g. enthalpy, number density, relaxation time that enters in the Eq.~\eqref{equ45ver1} can be obtained for the hadronic matter modeled by the hadron resonance gas model.

\section{Hadron resonance gas model}
\label{hrgmodels}
The ideal hadron resonance gas model (IHRG) model which is based on the Dashen, Ma, and Bernstein theorem~\cite{PhysRev.187.345,PhysRevD.10.694}, indicates that a dilute system of strongly interacting matter can be described by a gas of free hadrons and resonances. It should be emphasized that the strong nuclear force has attractive as well as repulsive parts. Both the long-range
attraction and the short-range repulsion are important for a consistent description of the strongly interacting matter~\cite{PhysRevC.88.024902}. The attractive part of the interaction is taken care of by the resonances. This free gas of hadrons and resonances constitutes the ideal hadron resonance gas model (IHRG) model. However, the repulsive nature of the nuclear force is not manifested by the presence of resonances~\cite{Andronic:2012ut}.  
The repulsive part of the strong nuclear force is incorporated
through the excluded-volume effects~\cite{Rischke:1991ke,Andronic:2012ut}. The
excluded-volume HRG (EVHRG) model has been used in the hydrodynamical models of nucleus-nucleus collision~\cite{Hama:2004rr,PhysRevC.82.044904}, to study the correlation and fluctuation of conserved charges~\cite{Bhattacharyya:2013oya}, viscous coefficients of hadronic matter~\cite{Kadam:2015xsa}, etc.

The grand canonical partition function of an ideal hadron resonance gas (IHRG) model can be written as~\cite{Braun-Munzinger:1994ewq,Braun-Munzinger:2003pwq}
\begin{align}
    \ln Z^{id}=\sum_a\ln Z_a^{id},
\end{align}
where the sum $`a'$ is over all the hadrons and resonances, $`id'$ indicates the non interacting hadron resonance gas. The partition function of the $`a'$th species, 
\begin{align}
\ln Z_a^{id} = \pm \frac{Vg_a}{2\pi^2}\int_0^{\infty}dp~p^2\ln[1\pm\exp(-\beta(E_a-\mu_a))].
\end{align}
Here $V$ is the volume of the system, $g_a$ is the degeneracy factor, $E_a$ is the single-particle energy, $\mu_a=\sum_q q_a\mu_q\equiv B_a\mu_B+S_a\mu_S+Q_a\mu_Q$ is the chemical potential, $B_a, S_a, Q_a$ are respectively the baryon number, strangeness number and electric charge. The $(+)$ and $(-)$ sign corresponds to fermions and bosons respectively. Note that we present our results only for the Boltzmann limit. Due to the conservation of different quantum numbers like baryon
number, charge, and strangeness various chemical potentials, e.g. $\mu_B$, $\mu_S$, $\mu_Q$ are not independent. However, for simplicity, we assume $\mu_S=0=\mu_Q$.  Once the partition function is known all the thermodynamic quantities of the system, e.g. pressure ($P^{id}$), net number density associated with conserved charges ($n_q^{id}$), energy density ($\varepsilon^{id}$) can be obtained using various thermodynamic relations~\cite{Braun-Munzinger:1994ewq,Braun-Munzinger:2003pwq,Andronic:2012ut}. 

Let us now discuss the excluded volume HRG (EVHRG) model. To consider the short-range repulsive
hadron-hadron interaction in the EVHRG model the geometrical size of the hadrons are explicitly incorporated as the excluded volume correction~\cite{Hagedorn:1980kb,Rischke:1991ke,Cleymans:1992jz,PhysRevC.56.2210}. A thermodynamically consistent excluded volume HRG model the pressure can be written as, 
\begin{align}
    P^{ex}(T,\mu_1,\mu_2,...)=\sum_a P_a^{id}(T,\tilde{\mu}_1,\tilde{\mu}_2,..),
\end{align}
where the chemical potential of the $`a'$th particle is, 
\begin{align}
    \tilde{\mu}_a=\mu_a-V^{ex}_{a}P^{ex}(T,\mu_1,\mu_2,..)
\end{align}
where $V^{ex}_a = (16/3)\pi R_a^3$ is the excluded volume for the $`a'$ th hadron. In an iterative procedure one can get the total pressure $P^{ex}(T,\mu_1,\mu_2,...)$. Other thermodynamic quantities which can be obtained from the pressure using appropriate thermodynamic relations can be given as, 
\begin{align}
    & n^{ex}=\sum_a\frac{n^{id}_a(T,\tilde{\mu}_a)}{1+\sum_b V^{ex}_b n_b^{id}(T,\tilde{\mu}_b)},\\
    & s^{ex}=\sum_a\frac{s^{id}_a(T,\tilde{\mu}_a)}{1+\sum_b V^{ex}_b n_b^{id}(T,\tilde{\mu}_b)},\\
    & \varepsilon^{ex}=\sum_a\frac{\varepsilon^{id}_a(T,\tilde{\mu}_a)}{1+\sum_b V^{ex}_b n_b^{id}(T,\tilde{\mu}_b)}.
\end{align}
In general, the hardcore radius of mesons and the baryons can be considered to be different. However, the thermodynamics of the EVHRG model is not strongly dependent on different values of the hardcore radius of the mesons and baryons ~\cite{Bhattacharyya:2013oya}. Therefore we consider here same hardcore radius for the mesons and baryons.
The remaining unknown quantity in the expression of $\kappa_{qq^{\prime}}$ is the relaxation time. In general relaxation time depends on the energy and the momentum of the particles involved in the scattering process. However for simplicity, one can integrate energy-dependent relaxation time over equilibrium distribution functions to get the thermal averaged relaxation time~\cite{PhysRevC.92.035203,Das:2021qii}. Without going into the details of calculating the energy averaged relaxation time we give here the important equations only. The thermal averaged relaxation time $\tau_a$ of the hadron species $``a"$ in terms of the scattering cross-section can be expressed as, 
\begin{align}
    \tau_a^{-1}=\sum_b n_b\langle \sigma_{ab}v_{ab}\rangle,
    \label{equ46ver1}
\end{align}
here $n_b$ denotes the number density of particle $``b"$ and $\langle \sigma_{ab}v_{ab}\rangle$ represents thermal averaged cross section. In Eq.~\eqref{equ46ver1} the sum is over all the hadrons and its resonances. For hard sphere scattering in the Boltzmann approximation the thermal averaged cross section can be expressed as~\cite{PhysRevC.92.035203},
\begin{align}
    \langle \sigma_{ab}v_{ab}\rangle & =\frac{\sigma}{8Tm_a^2m_b^2K_2(m_a/T)K_2(m_b/T)}\nonumber\\
    &\times\int_{(m_a+m_b)^2}^{\infty}ds\times \frac{[s-(m_a-m_b)^2]}{\sqrt{s}}\nonumber\\
    &\times [s-(m_a+m_b)^2]K_1(\sqrt{s}/T),
\end{align}
here the hard sphere scattering cross section can be expressed as $\sigma=4\pi R^2$. $R$ is the radius of hadrons. Once the thermal averaged relaxation time is known for each hadron species, different components of the diffusion matrix can be obtained using Eq.~\eqref{equ45ver1}.  

\begin{figure}[]
	\includegraphics[scale=0.5]{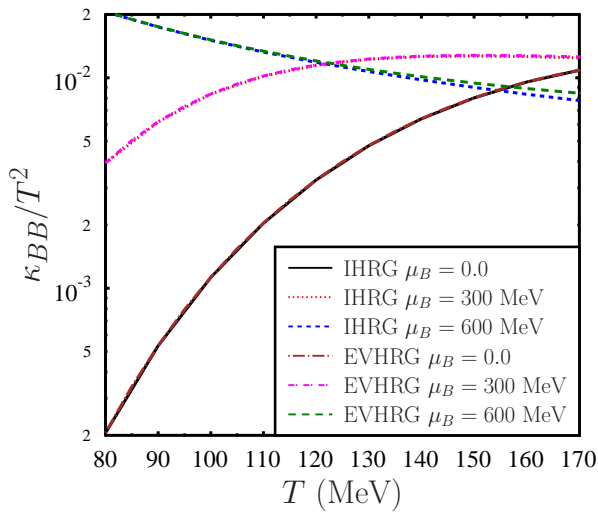}
	\caption{Variation of dimensionless ratio $\kappa_{BB}/T^2$ with temperature ($T$) and the baryon chemical potential ($\mu_B$). Here we have assumed $\mu_Q=0=\mu_S$. Physically $\kappa_{BB}$ is the measure of the baryon number current generation due to gradient in $\beta\mu_B$. Among all the hadrons baryonic contribution is dominant over mesonic contribution in $\kappa_{BB}/T^2$. Estimated values of $\kappa_{BB}/T^2$ are almost similar in IHRG and EVHRG models. Different curves are overlapping.}
	\label{kappabb}
\end{figure}

\begin{figure}[]
	\includegraphics[scale=0.5]{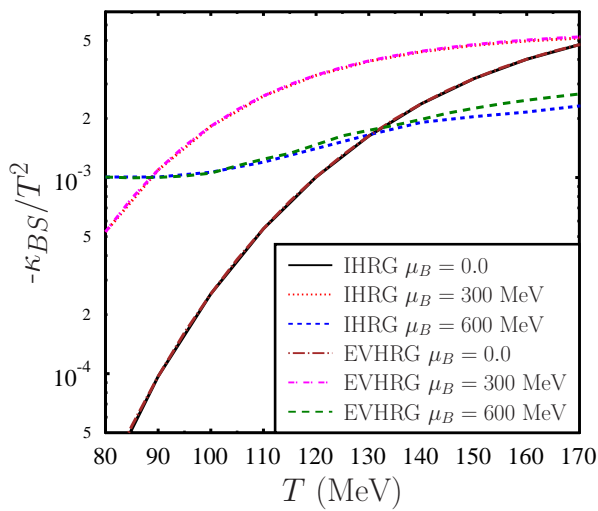}
	\caption{Variation of dimensionless ratio $-\kappa_{BS}/T^2$ with temperature ($T$) and the baryon chemical potential ($\mu_B$) for $\mu_Q=0=\mu_S$. For the range of thermodynamic parameters considered here $\kappa_{BS}$ is always negative. Physically $\kappa_{BS}$ is the measure of the diffusive coupling between the baryon number and the strangeness number. Among all the hadrons baryons contribute dominantly in $\kappa_{BS}/T^2$. Values of $\kappa_{BS}/T^2$ as obtained in IHRG and EVHRG models are very similar with overlapping curves.}
	\label{kappabs}
\end{figure}

\begin{figure}[]
	\includegraphics[scale=0.5]{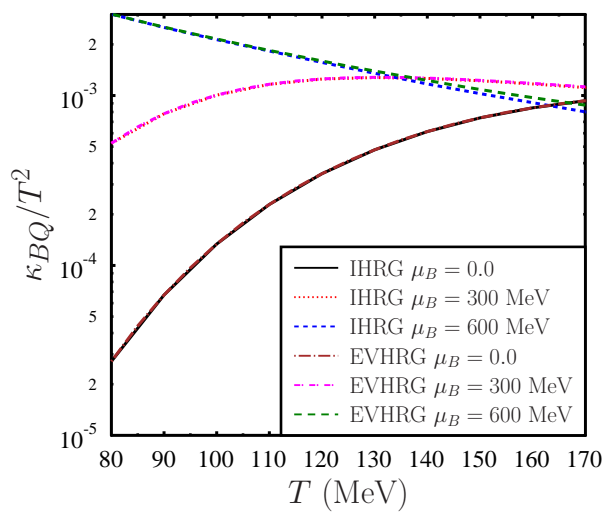}
	\caption{Variation of dimensionless ratio $\kappa_{BQ}/T^2$ with temperature ($T$) and the baryon chemical potential ($\mu_B$) for $\mu_Q=0=\mu_S$. $\kappa_{BQ}$ is the measure of the diffusive coupling between the baryon and electric charge. Among all the hadrons leading contribution  in $\kappa_{BQ}/T^2$ comes from the baryons. Estimated values of $\kappa_{BQ}/T^2$ in IHRG and EVHRG models are very similar with overlapping curves.}
	\label{kappabq}
\end{figure}

\begin{figure}[t]
	\includegraphics[scale=0.5]{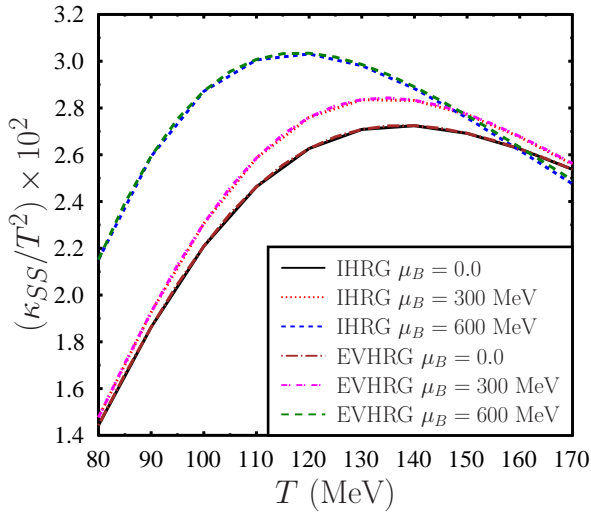}
	\caption{Variation of diagonal component $\kappa_{SS}/T^2$ with temperature ($T$) and the baryon chemical potential ($\mu_B$). Here also we consider the scenario with $\mu_Q=0=\mu_S$. $\kappa_{SS}$ is the measure of the strangeness number current due to the gradient in $\beta\mu_S$. Since the lightest strange hadrons are mesons, therefore mesonic contribution is dominant in $\kappa_{SS}/T^2$ with respect to the baryonic contribution. Estimated values of $\kappa_{SS}/T^2$ in IHRG and EVHRG models are very similar with overlapping curves.}
	\label{kappass}
\end{figure}

\begin{figure}[t]
	\includegraphics[scale=0.5]{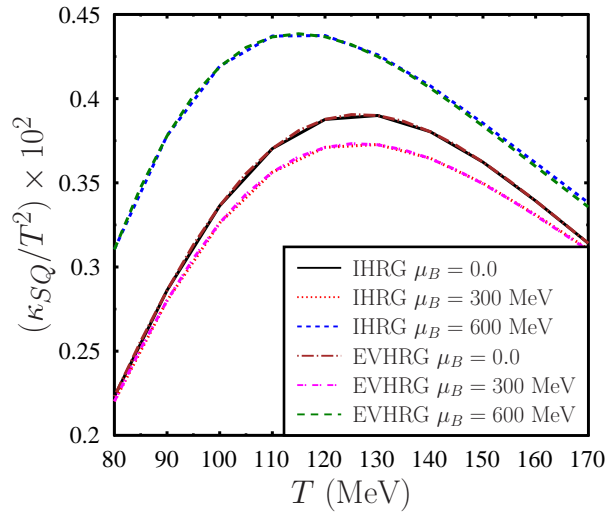}
	\caption{Variation of dimensionless ratio $\kappa_{SQ}/T^2$ with temperature ($T$) and the baryon chemical potential ($\mu_B$) for $\mu_Q=0=\mu_S$. $\kappa_{SQ}$ is the measure of the diffusive coupling between the strangeness charge and electric charge. Mesons contribute predominantly in $\kappa_{SQ}/T^2$. Different curves for IHRG and EVHRG are overlapping.}
	\label{kappasq}
\end{figure}

\begin{figure}[t]
	\includegraphics[scale=0.5]{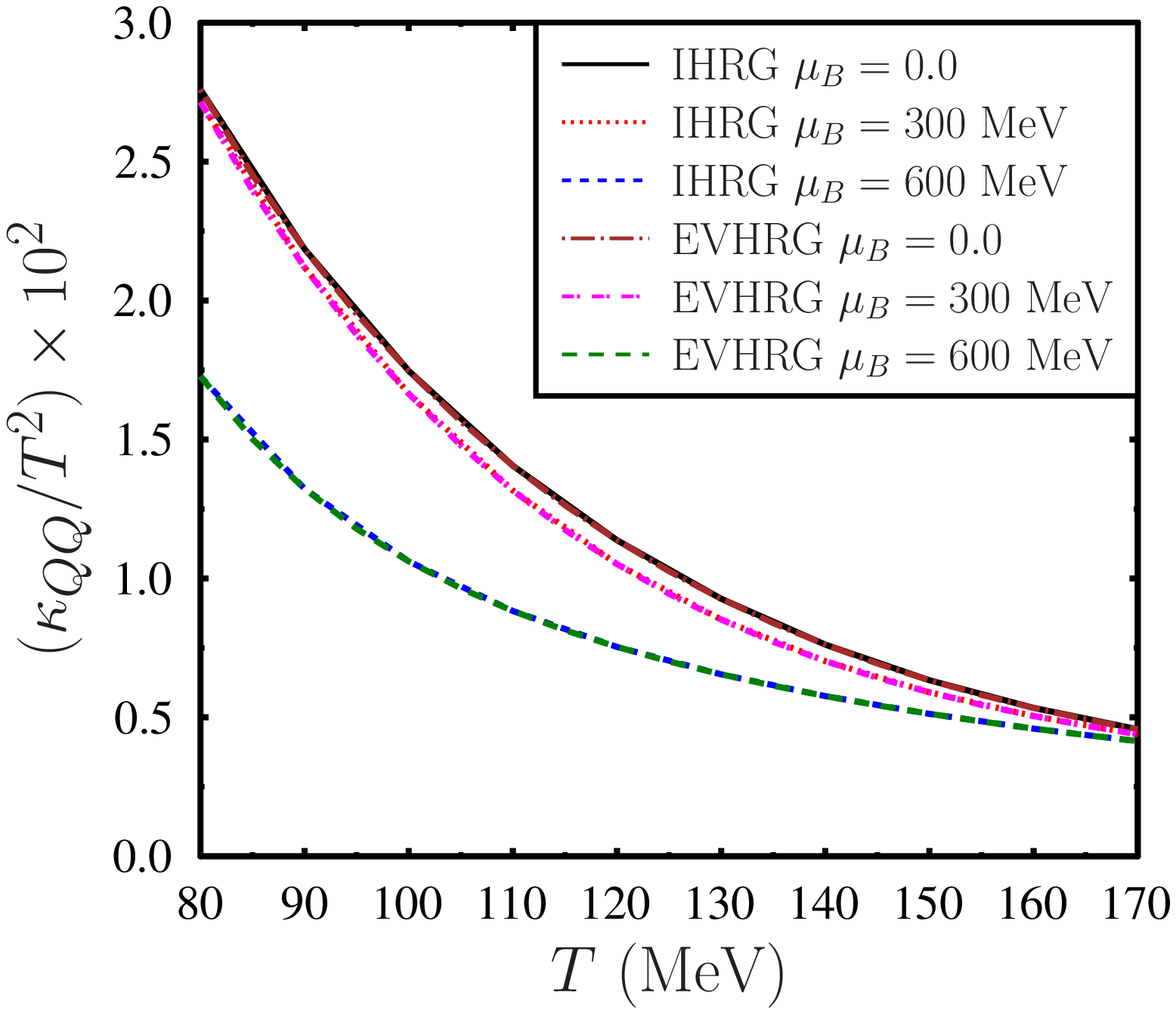}
	\caption{Variation of diagonal component $\kappa_{QQ}/T^2$ with temperature ($T$) and the baryon chemical potential ($\mu_B$).  $\kappa_{QQ}$ is the measure of the electrical current due to the gradient in $\beta\mu_Q$. For the hadron resonance gas leading contribution to $\kappa_{QQ}/T^2$ comes from the mesons with respect to the hadrons. Similar to other elements of the diffusion matrix, $\kappa_{QQ}/T^2$ also has similar values in the IHRG and EVHRG model.}
	\label{kappaqq}
\end{figure}

\section{results and discussions}
\label{results}
We present the results for the diffusion matrix elements $\kappa_{qq^{\prime}}$ with temperature and baryon chemical potential. For simplicity, we assume that electric chemical potential and the strangeness chemical potential is zero, i.e. $\mu_Q=0=\mu_S$. For HRG, we consider all the hadrons and their resonances up to a mass cutoff $\Lambda=2.6$ GeV, as is listed in Ref.~\cite{ParticleDataGroup:2008zun}. Also for a detailed list of hadrons and their resonances, we refer to
Appendix A of Ref.~\cite{PhysRevC.90.024915}. These apart radii of the hard spheres also enter in the calculation of relaxation time. We consider a uniform radius $R = 0.5$ fm for all the hadrons~\cite{PhysRevC.92.035203,Braun-Munzinger:1999hun}.

In Fig.~\eqref{kappabb} we show the variation of the diagonal component of the diffusion matrix associated with the baryon number current, i.e. $\kappa_{BB}$ with temperature and baryon chemical potential. For $\mu_B=0$ and 300 MeV the dimensionless quantity $\kappa_{BB}/T^2$ increases with temperature. However for $\mu_B=600$ MeV, $\kappa_{BB}/T^2$ decreases with temperature. Since here we consider $\mu_Q=0=\mu_S$, for vanishing value of baryon chemical potential the variation of $\kappa_{BB}/T^2$ can be understood in the following manner. For $\mu_B=\mu_Q=\mu_S=0$, the net baryon number density vanishes, i.e. $n_B=0$. Note that with temperature the relaxation time of hadrons decreases. On the other hand with temperature the contribution coming from the distribution function increase due to the increase in the baryon numbers. This increase in the distribution function wins over the decrease in the relaxation time, giving rise to an increasing behavior of $\kappa_{BB}/T^2$ with temperature. However for $\mu_B\neq 0$, in the expression of $\kappa_{BB}$, $n_B/\omega$ increases with temperature~\cite{Das:2019pqd} along with the decreases in the relaxation time with temperature. For a sufficiently large value of the baryon chemical potential, this decrease in the relaxation time is predominant giving rise to a decreasing behavior of $\kappa_{BB}/T^2$ with temperature for $\mu_B=600$ MeV.

Further in the relatively low-temperature range $\kappa_{BB}/T^2$ increases with $\mu_B$. Note that for the hadron resonance gas with $\mu_B$, $n_B/\omega$ increases~\cite{Das:2019pqd} along with the increase in the distribution function. Again the relaxation time of various hadron species decreases with the baryon chemical potential. Such an increase in $n_B/\omega$ and distribution function dominates over the decreases in the relaxation time giving rise to an overall increasing trend. However, for a sufficiently high-temperature range, the variation of $\kappa_{BB}/T^2$ with the baryon chemical potential is nonmonotonic. This nonmonotonic variation is predominantly due to the decrease in the relaxation time with temperature and baryon chemical potential.    
Note that results for $\kappa_{BB}/T^2$ as obtained in the IHRG, as well as EVHRG, are almost similar. This is a generic feature for other components of the diffusion matrix. Various thermodynamic quantities, e.g. pressure, energy density, and number density are different in EVHRG as compared to the IHRG model. But the thermodynamic quantity that enters into the expression of $\kappa_{qq}$ is $n_q/\omega$. $n_q/\omega$ does not change significantly in the EVHRG as compared to the IHRG, giving rise to an almost similar variation of the diffusion matrix elements with temperature and baryon chemical potential.  

In Fig.~\eqref{kappabs} we show the variation of $-\kappa_{BS}/T^2$ with temperature ($T$) and the baryon chemical potential ($\mu_B$). Physically a nonvanishing value of $\kappa_{BS}$ indicates the generation of baryon current due to the gradient in the number density of hadrons containing strangeness quantum number. Further note that $\kappa_{BS}$ is negative for the range of temperature and baryon chemical potential considered here. For $\mu_B=0$ this can be understood in the following way. For $\mu_B=\mu_S=\mu_Q=0$, $n_B=n_S=0$ and $\kappa_{BS}$ is proportional to the (baryon number $\times$ strangeness number). Now in the baryon sector, baryon number and strangeness number of various baryons are opposite in sign giving rise to a negative value of $\kappa_{BS}$. Also for finite $\mu_B$ baryon contributes dominantly in $\kappa_{BS}$ due to an increase in the number density of baryons. The opposite sign of the baryon number and the strangeness number for baryons give rise to an overall negative sign to $\kappa_{BS}$.  Similar to $\kappa_{BB}/T^2$, the off-diagonal component $-\kappa_{BS}/T^2$ also shows strong temperature dependence for $\mu_B=0, 300$ MeV. Such a behaviour is again predominantly due to increase in the equilibrium distribution function.  The nonmonotonic variation of $-\kappa_{BS}/T^2$ with $\mu_B$ is again due to various $\mu_B$ dependent factors, e.g. relaxation time, $n_B/\omega$, $n_S/\omega$ and the equilibrium distribution function.  

In Fig.~\eqref{kappabq} we show the variation $\kappa_{BQ}/T^2$ with temperature ($T$) and the baryon chemical potential ($\mu_B$). Similar to $\kappa_{BS}$ a nonvanishing value of $\kappa_{BQ}$ indicates gradient in the baryon number density can give rise to electric current. Further note that unlike $\kappa_{BS}$, $\kappa_{BQ}$ is positive for the range of temperature and baryon chemical potential considered here. Such behavior is easy to understand for $\mu_B=0$ case. As discussed earlier for  $\mu_B=\mu_S=\mu_Q=0$, $n_B=n_S=0$ and $\kappa_{BQ}$ is proportional to the (baryon number $\times$ electric charge). Now in the baryon sector for the lightest baryon and antibaryons, baryon number and electric charge has the same sign giving rise to a positive value of $\kappa_{BQ}$. Similarly one can also explain the overall positive sign for $\kappa_{BQ}/T^2$at finite $\mu_B$.
Variation of $\kappa_{BQ}/T^2$ with temperature and baryon chemical potential is very similar to $\kappa_{BB}/T^2$. Such variation of $\kappa_{BQ}/T^2$ with temperature and baryon chemical can be qualitatively understood by looking into the behavior of relaxation time, $n_B/\omega$, $n_Q/\omega$ and the equilibrium distribution function with $T$ and $\mu_B$.

In Figs.~\eqref{kappass} and \eqref{kappasq} we present the results for the diffusion coefficients associated with the strangeness current originated due to gradient in $\beta\mu_S$ and $\beta\mu_Q$ respectively. According to Eq.~\eqref{equ45ver1} the diagonal component $\kappa_{SS}/T^2$ is always positive. The off-diagonal component $\kappa_{SQ}/T^2$ is also positive for the range of temperature and baryon chemical potential considered here. Both $\kappa_{SS}/T^2$ and $\kappa_{SQ}/T^2$ shows nonmonotonic variation with temperature and baryon chemical potential. Such nonmontonic variation of $\kappa_{SS}/T^2$ and $\kappa_{SQ}/T^2$ is rather convoluted as various factors, e.g. the relaxation time, distribution function, $n_S/\omega$ and $n_Q/\omega$ depends upon temperature and baryon chemical potential.  Although the behaviour of $\kappa_{BB}/T^2$, $\kappa_{BS}/T^2$ and $\kappa_{BQ}/T^2$ as presented in Figs.~\eqref{kappabb}, \eqref{kappabs} and \eqref{kappabq} respectively, are similar to the results presented in the Refs.~\cite{Fotakis:2019nbq,Greif:2017byw}, the nonmonotonic variation of $\kappa_{SS}/T^2$ and $\kappa_{SQ}/T^2$ is different from the results presented in the Refs.~\cite{Fotakis:2019nbq,Greif:2017byw}. 

Finally, in Fig.~\eqref{kappaqq} we show the variation of $\kappa_{QQ}/T^2$ with $T$ and $\mu_B$. As argued in Refs.~\cite{Fotakis:2019nbq,Greif:2017byw}, here also for $\mu_B=0$, $\kappa_{QQ}/T^2\equiv\sigma_{el}/T$. $\sigma_{el}$ is the electrical conductivity of the medium in the kinetic theory approach
~\cite{Das:2019wjg}. Although for non vanishing values of baryon chemical potential  $\kappa_{QQ}/T^2\neq\sigma_{el}/T$, but the qualitative behaviour of $\kappa_{QQ}/T^2$ and $\sigma_{el}/T$ are similar even for $\mu_B\neq0$~\cite{Das:2019wjg}. Among all the other hadrons pions and protons contribute dominantly in $\sigma_{el}/T$ as well as in $\kappa_{QQ}/T^2$. As argued in Ref.~\cite{Das:2019wjg} among various temperature and baryon chemical potential dependent quantities, due to the decrease of the relaxation time of hadrons with $T$ and $\mu_B$, $\sigma_{el}/T$ or $\kappa_{QQ}/T^2$ also decreases. 

\section{conclusion}
\label{conclusion}
In the present investigation, we discuss the diffusion matrix associated with the various conserved quantities. Using the classical kinetic theory within the relaxation time approximation we obtained an analytical expression of the diffusion matrix ($\kappa_{qq^{\prime}}$). The diagonal components of the diffusion matrix are always positive, on the other hand, the off-diagonal components can be negative as well as positive. Using the hadron resonance gas model within the hard-sphere scattering approximation we estimated various elements of $\kappa_{qq^{\prime}}$ for the hadronic medium produced in heavy-ion collisions. Knowledge of the diffusion processes and the diffusion currents is very important in the context of the bulk evolution of the strongly interacting plasma, e.g. the cross-coupling between the diffusion currents can dynamically generate non-zero net strangeness, even if it is initially zero~\cite{Fotakis:2019nbq}. Further, the order of magnitude values of the off-diagonal components of the diffusion matrix is not at all negligible as compared to the diagonal elements. Therefore, it is of paramount importance to computing the full table of diffusion coefficients for consistent fluid dynamical simulations where the diffusion is also taken under consideration. 
In this paper, we have ignored the effect of the mean-field or medium modification on the constituent of the plasma. Such mean-field effects should be included in the kinetic theory description as it might be quite important across the QCD transition scale.  

\begin{acknowledgments}
The work of AD is supported by the Polish National Science Center Grant No. 2018/30/E/ST2/00432. AD would like to thank Guru Prakash Kadam for important discussions on the excluded volume hadron resonance gas model. 
\end{acknowledgments}

\appendix
\section{}
\label{appenA}
Let us start with the following integration for a single species, 
\begin{align}
    \mathcal{I}_1 & =\int \frac{d^3p}{(2\pi)^3}p^2 \exp(-\beta\sqrt{p^2+m^2})\nonumber\\
    & =\frac{m^5}{2\pi^2}\int_0^{\infty}dy~\cosh{y}~(\sinh{y})^4\exp(-\beta m \cosh{y})\nonumber\\
    & = \frac{m^5}{2\pi^2}\int_0^{\infty}dy~((\cosh{y})^5-2(\cosh{y})^3+\cosh{y})\nonumber\\
    &~~~~~~~~~~~~~~~~~~~~~~\times\exp(-\beta m \cosh{y})\nonumber\\
    & = \frac{m^5}{2\pi^2} \bigg(\mathcal{I}_{1a}-2\mathcal{I}_{1b}+\mathcal{I}_{1c}\bigg),
    \label{eqappen1}
\end{align}
here, 
\begin{align}
    & \mathcal{I}_{1a}=\int_0^{\infty}dy (\cosh{y})^5 \exp(-\beta m \cosh{y}), \\
    & \mathcal{I}_{1b}=\int_0^{\infty}dy (\cosh{y})^3 \exp(-\beta m \cosh{y}), \\
    & \mathcal{I}_{1c}=\int_0^{\infty}dy \cosh{y} \exp(-\beta m \cosh{y}).
\end{align}
Using the integral representation of the Modified Bessel Function of the second kind ($K_n(x)$) the integral it is easy to show that,
\begin{align}
     & \mathcal{I}_{1c}=K_1(\beta m),\label{eqappen5}\\
     & \mathcal{I}_{1b} = \frac{\partial^2 K_1(\beta m)}{\partial(\beta m)^2} = \frac{1}{\beta m}K_0(\beta m)+\frac{2}{(\beta m)^2}K_1(\beta m)\nonumber\\
     &~~~~~~~~~~~~~~~~~~~~~~~~~~~~~~~~~+K_1(\beta m)\nonumber\\
     &~~~~~~~ = \frac{1}{\beta m}K_2(\beta m)+K_1(\beta m),\label{eqappen6}\\
     & \mathcal{I}_{1a} = \frac{\partial^4 K_1(\beta m)}{\partial(\beta m)^4} \nonumber\\
     & =\frac{12}{(\beta m)^3}K_0(\beta m)+\frac{2}{\beta m} K_0(\beta m)+\frac{24}{(\beta m)^4}K_1(\beta m)\nonumber\\
     & ~~~~~~~\frac{7}{(\beta m)^2}K_1(\beta m)+K_1(\beta m)\nonumber\\
     & = \frac{3}{(\beta m)^2}K_3(\beta m)+\frac{2}{\beta m}K_2(\beta m)+K_1(\beta m).
     \label{eqappen7}
\end{align}
Eqs.~\eqref{eqappen5}-\eqref{eqappen7} allows us to write the integral $\mathcal{I}_1$ as, 
\begin{align}
    \mathcal{I}_1=\frac{3m^3T^2}{2\pi^2}K_3(\beta m). 
\end{align}
Now the energy density and pressure of a single particle species of mass $m$ at finite temperature $T$ can be expressed as~\cite{Florkowski:2014sfa}, 
\begin{align}
\varepsilon = \frac{m^2 T}{2\pi^2}\bigg(3TK_2(\beta m)+m K_1(\beta m)\bigg),
\end{align}
and, 
\begin{align}
    P = \frac{m^2T^2}{2\pi^2}K_2(\beta m).
\end{align}
Therefore, the enthalpy ($\omega$) can be expressed as, 
\begin{align}
    \omega & = \varepsilon+P\nonumber\\
    & = \frac{4 m^2 T^2}{2\pi^2}K_2(\beta m)+\frac{m^3 T}{2\pi^2}K_1(\beta m)\nonumber\\
    & = \frac{m^3 T}{2\pi^2}K_3(\beta m) = \frac{1}{3 T}\mathcal{I}_1.
\end{align}
The above expression can be generalized to multiple particle species. 
\section{}
\label{appenB}
   We give here a short derivation of Eq.(\ref{equ40ver1}).
 Without loss of generality,let us consider only the baryon number conservation and a single baryon and it's antibaryon species. In this case the integral, 
\begin{align}
    \mathcal{I}_2 & =\sum_a q_a \int \frac{d^3p_a}{(2\pi)^3}\frac{p_a^2}{3E_a}\exp(-\beta \sqrt{p_a^2+m_a^2})\exp(\beta q_a\mu_q)\nonumber\\
    & = \int \frac{d^3p}{(2\pi^3)}\frac{p^2}{3E} \exp(-\beta \sqrt{p^2+m^2})2\sinh(\beta\mu_B) \nonumber\\
    & = 2\sinh(\beta\mu_B)\frac{m^4}{6\pi^2}\int_0^{\infty}dy (\sinh{y})^4\exp(-\beta m\cosh{y})\nonumber\\
    & = 2\sinh(\beta\mu_B)\frac{m^4}{6\pi^2} (\mathcal{I}_{2a}-2\mathcal{I}_{2b}+\mathcal{I}_{2c}),
\end{align}
here, 
\begin{align}
    & \mathcal{I}_{2a}=\int_0^{\infty}dy (\cosh{y})^4 \exp(-\beta m \cosh{y}), \\
    & \mathcal{I}_{2b}=\int_0^{\infty}dy (\cosh{y})^2 \exp(-\beta m \cosh{y}), \\
    & \mathcal{I}_{1c}=\int_0^{\infty}dy  \exp(-\beta m \cosh{y}).
\end{align}
Using the integral representation of the Modified Bessel Function of the second kind ($K_n(x)$) it can be shown that, 
\begin{align}
    & \mathcal{I}_{2c} = K_0(\beta m), \\
    & \mathcal{I}_{2b} 
    = K_0(\beta m)+\frac{1}{\beta m}K_1(\beta m),\\
    & \mathcal{I}_{2a} 
    = \frac{3}{(\beta m)^2}K_2(\beta m)+K_2(\beta m).
\end{align}
Therefore, 
\begin{align}
    \mathcal{I}_2 
     =  2\sinh(\beta\mu_B) \frac{m^2T^2}{2\pi^2}K_2(\beta m).
\end{align}
Net baryon number density can be expressed as, 
\begin{align}
    n_B & = 2\sinh{\beta\mu_B}\int\frac{d^3p}{(2\pi)^3}\exp(-\beta m\sqrt{p^2+m^2})\nonumber\\
    & = 2\sinh{\beta\mu_B} \frac{m^2 T}{2\pi^2}K_2(\beta m).
\end{align}
Therefore, 
\begin{align}
    \mathcal{I}_2=n_BT.
\end{align}
The above expression can be easily generalized to include other conserved charges. 

\bibliography{fluctuationRef.bib}{}

\providecommand{\href}[2]{#2}\begingroup\raggedright\begin{thebibliography}{100}

\bibitem{Romatschke:2017ejr}
P.~Romatschke and U.~Romatschke,
  \href{http://dx.doi.org/10.1017/9781108651998}{{\em {Relativistic Fluid
  Dynamics In and Out of Equilibrium}}}.
\newblock Cambridge Monographs on Mathematical Physics. Cambridge University
  Press, 5, 2019.
\newblock \href{http://arxiv.org/abs/1712.05815}{{\ttfamily arXiv:1712.05815
  [nucl-th]}}.

\bibitem{Florkowski:2010zz}
W.~Florkowski, {\em {Phenomenology of Ultra-Relativistic Heavy-Ion
  Collisions}}.
\newblock 3, 2010.

\bibitem{Gale:2013da}
C.~Gale, S.~Jeon, and B.~Schenke, ``{Hydrodynamic Modeling of Heavy-Ion
  Collisions},'' \href{http://dx.doi.org/10.1142/S0217751X13400113}{{\em Int.
  J. Mod. Phys. A} {\bfseries 28} (2013) 1340011},
  \href{http://arxiv.org/abs/1301.5893}{{\ttfamily arXiv:1301.5893 [nucl-th]}}.

\bibitem{Jeon:2015dfa}
S.~Jeon and U.~Heinz, ``{Introduction to Hydrodynamics},''
  \href{http://dx.doi.org/10.1142/S0218301315300106}{{\em Int. J. Mod. Phys. E}
  {\bfseries 24} no.~10, (2015) 1530010},
  \href{http://arxiv.org/abs/1503.03931}{{\ttfamily arXiv:1503.03931
  [hep-ph]}}.

\bibitem{Jaiswal:2016hex}
A.~Jaiswal and V.~Roy, ``{Relativistic hydrodynamics in heavy-ion collisions:
  general aspects and recent developments},''
  \href{http://dx.doi.org/10.1155/2016/9623034}{{\em Adv. High Energy Phys.}
  {\bfseries 2016} (2016) 9623034},
  \href{http://arxiv.org/abs/1605.08694}{{\ttfamily arXiv:1605.08694
  [nucl-th]}}.

\bibitem{Heinz:2013th}
U.~Heinz and R.~Snellings, ``{Collective flow and viscosity in relativistic
  heavy-ion collisions},''
  \href{http://dx.doi.org/10.1146/annurev-nucl-102212-170540}{{\em Ann. Rev.
  Nucl. Part. Sci.} {\bfseries 63} (2013) 123--151},
  \href{http://arxiv.org/abs/1301.2826}{{\ttfamily arXiv:1301.2826 [nucl-th]}}.

\bibitem{Kovtun:2004de}
P.~Kovtun, D.~T. Son, and A.~O. Starinets, ``{Viscosity in strongly interacting
  quantum field theories from black hole physics},''
  \href{http://dx.doi.org/10.1103/PhysRevLett.94.111601}{{\em Phys. Rev. Lett.}
  {\bfseries 94} (2005) 111601},
  \href{http://arxiv.org/abs/hep-th/0405231}{{\ttfamily arXiv:hep-th/0405231}}.

\bibitem{Gavin:1985ph}
S.~Gavin, ``{TRANSPORT COEFFICIENTS IN ULTRARELATIVISTIC HEAVY ION
  COLLISIONS},'' \href{http://dx.doi.org/10.1016/0375-9474(85)90190-3}{{\em
  Nucl. Phys. A} {\bfseries 435} (1985) 826--843}.

\bibitem{Hosoya:1983xm}
A.~Hosoya and K.~Kajantie, ``{Transport Coefficients of QCD Matter},''
  \href{http://dx.doi.org/10.1016/0550-3213(85)90499-7}{{\em Nucl. Phys. B}
  {\bfseries 250} (1985) 666--688}.

\bibitem{Dobado:2012zf}
A.~Dobado and J.~M. Torres-Rincon, ``{Bulk viscosity and the phase transition
  of the linear sigma model},''
  \href{http://dx.doi.org/10.1103/PhysRevD.86.074021}{{\em Phys. Rev. D}
  {\bfseries 86} (2012) 074021},
  \href{http://arxiv.org/abs/1206.1261}{{\ttfamily arXiv:1206.1261 [hep-ph]}}.

\bibitem{Sasaki:2008fg}
C.~Sasaki and K.~Redlich, ``{Bulk viscosity in quasi particle models},''
  \href{http://dx.doi.org/10.1103/PhysRevC.79.055207}{{\em Phys. Rev. C}
  {\bfseries 79} (2009) 055207},
  \href{http://arxiv.org/abs/0806.4745}{{\ttfamily arXiv:0806.4745 [hep-ph]}}.

\bibitem{Sasaki:2008um}
C.~Sasaki and K.~Redlich, ``{Transport coefficients near chiral phase
  transition},'' \href{http://dx.doi.org/10.1016/j.nuclphysa.2009.11.005}{{\em
  Nucl. Phys. A} {\bfseries 832} (2010) 62--75},
  \href{http://arxiv.org/abs/0811.4708}{{\ttfamily arXiv:0811.4708 [hep-ph]}}.

\bibitem{Karsch:2007jc}
F.~Karsch, D.~Kharzeev, and K.~Tuchin, ``{Universal properties of bulk
  viscosity near the QCD phase transition},''
  \href{http://dx.doi.org/10.1016/j.physletb.2008.01.080}{{\em Phys. Lett. B}
  {\bfseries 663} (2008) 217--221},
  \href{http://arxiv.org/abs/0711.0914}{{\ttfamily arXiv:0711.0914 [hep-ph]}}.

\bibitem{Finazzo:2014cna}
S.~I. Finazzo, R.~Rougemont, H.~Marrochio, and J.~Noronha, ``{Hydrodynamic
  transport coefficients for the non-conformal quark-gluon plasma from
  holography},'' \href{http://dx.doi.org/10.1007/JHEP02(2015)051}{{\em JHEP}
  {\bfseries 02} (2015) 051}, \href{http://arxiv.org/abs/1412.2968}{{\ttfamily
  arXiv:1412.2968 [hep-ph]}}.

\bibitem{Wiranata:2009cz}
A.~Wiranata and M.~Prakash, ``{Bulk Viscosity of Interacting Hadrons},''
  \href{http://dx.doi.org/10.1016/j.nuclphysa.2009.09.023}{{\em Nucl. Phys. A}
  {\bfseries 830} (2009) 219C--222C},
  \href{http://arxiv.org/abs/0906.5592}{{\ttfamily arXiv:0906.5592 [nucl-th]}}.

\bibitem{Jeon:1995zm}
S.~Jeon and L.~G. Yaffe, ``{From quantum field theory to hydrodynamics:
  Transport coefficients and effective kinetic theory},''
  \href{http://dx.doi.org/10.1103/PhysRevD.53.5799}{{\em Phys. Rev. D}
  {\bfseries 53} (1996) 5799--5809},
  \href{http://arxiv.org/abs/hep-ph/9512263}{{\ttfamily arXiv:hep-ph/9512263}}.

\bibitem{Noronha-Hostler:2013gga}
J.~Noronha-Hostler, G.~S. Denicol, J.~Noronha, R.~P.~G. Andrade, and F.~Grassi,
  ``{Bulk Viscosity Effects in Event-by-Event Relativistic Hydrodynamics},''
  \href{http://dx.doi.org/10.1103/PhysRevC.88.044916}{{\em Phys. Rev. C}
  {\bfseries 88} no.~4, (2013) 044916},
  \href{http://arxiv.org/abs/1305.1981}{{\ttfamily arXiv:1305.1981 [nucl-th]}}.

\bibitem{Ryu:2015vwa}
S.~Ryu, J.~F. Paquet, C.~Shen, G.~S. Denicol, B.~Schenke, S.~Jeon, and C.~Gale,
  ``{Importance of the Bulk Viscosity of QCD in Ultrarelativistic Heavy-Ion
  Collisions},'' \href{http://dx.doi.org/10.1103/PhysRevLett.115.132301}{{\em
  Phys. Rev. Lett.} {\bfseries 115} no.~13, (2015) 132301},
  \href{http://arxiv.org/abs/1502.01675}{{\ttfamily arXiv:1502.01675
  [nucl-th]}}.

\bibitem{Ryu:2017qzn}
S.~Ryu, J.-F. Paquet, C.~Shen, G.~Denicol, B.~Schenke, S.~Jeon, and C.~Gale,
  ``{Effects of bulk viscosity and hadronic rescattering in heavy ion
  collisions at energies available at the BNL Relativistic Heavy Ion Collider
  and at the CERN Large Hadron Collider},''
  \href{http://dx.doi.org/10.1103/PhysRevC.97.034910}{{\em Phys. Rev. C}
  {\bfseries 97} no.~3, (2018) 034910},
  \href{http://arxiv.org/abs/1704.04216}{{\ttfamily arXiv:1704.04216
  [nucl-th]}}.

\bibitem{Vujanovic:2019yih}
G.~Vujanovic, J.-F. Paquet, C.~Shen, G.~S. Denicol, S.~Jeon, C.~Gale, and
  U.~Heinz, ``{Exploring the influence of bulk viscosity of QCD on dilepton
  tomography},'' \href{http://dx.doi.org/10.1103/PhysRevC.101.044904}{{\em
  Phys. Rev. C} {\bfseries 101} (2020) 044904},
  \href{http://arxiv.org/abs/1903.05078}{{\ttfamily arXiv:1903.05078
  [nucl-th]}}.

\bibitem{Tuchin:2010gx}
K.~Tuchin, ``{Photon decay in strong magnetic field in heavy-ion collisions},''
  \href{http://dx.doi.org/10.1103/PhysRevC.83.017901}{{\em Phys. Rev. C}
  {\bfseries 83} (2011) 017901},
  \href{http://arxiv.org/abs/1008.1604}{{\ttfamily arXiv:1008.1604 [nucl-th]}}.

\bibitem{Tuchin:2010vs}
K.~Tuchin, ``{Synchrotron radiation by fast fermions in heavy-ion
  collisions},'' \href{http://dx.doi.org/10.1103/PhysRevC.83.039903}{{\em Phys.
  Rev. C} {\bfseries 82} (2010) 034904},
  \href{http://arxiv.org/abs/1006.3051}{{\ttfamily arXiv:1006.3051 [nucl-th]}}.
  [Erratum: Phys.Rev.C 83, 039903 (2011)].

\bibitem{Inghirami:2016iru}
G.~Inghirami, L.~Del~Zanna, A.~Beraudo, M.~H. Moghaddam, F.~Becattini, and
  M.~Bleicher, ``{Numerical magneto-hydrodynamics for relativistic nuclear
  collisions},'' \href{http://dx.doi.org/10.1140/epjc/s10052-016-4516-8}{{\em
  Eur. Phys. J. C} {\bfseries 76} no.~12, (2016) 659},
  \href{http://arxiv.org/abs/1609.03042}{{\ttfamily arXiv:1609.03042
  [hep-ph]}}.

\bibitem{Das:2017qfi}
A.~Das, S.~S. Dave, P.~S. Saumia, and A.~M. Srivastava, ``{Effects of magnetic
  field on plasma evolution in relativistic heavy-ion collisions},''
  \href{http://dx.doi.org/10.1103/PhysRevC.96.034902}{{\em Phys. Rev. C}
  {\bfseries 96} no.~3, (2017) 034902},
  \href{http://arxiv.org/abs/1703.08162}{{\ttfamily arXiv:1703.08162
  [hep-ph]}}.

\bibitem{Greif:2016skc}
M.~Greif, C.~Greiner, and G.~S. Denicol, ``{Electric conductivity of a hot
  hadron gas from a kinetic approach},''
  \href{http://dx.doi.org/10.1103/PhysRevD.93.096012}{{\em Phys. Rev. D}
  {\bfseries 93} no.~9, (2016) 096012},
  \href{http://arxiv.org/abs/1602.05085}{{\ttfamily arXiv:1602.05085
  [nucl-th]}}. [Erratum: Phys.Rev.D 96, 059902 (2017)].

\bibitem{Greif:2014oia}
M.~Greif, I.~Bouras, C.~Greiner, and Z.~Xu, ``{Electric conductivity of the
  quark-gluon plasma investigated using a perturbative QCD based parton
  cascade},'' \href{http://dx.doi.org/10.1103/PhysRevD.90.094014}{{\em Phys.
  Rev. D} {\bfseries 90} no.~9, (2014) 094014},
  \href{http://arxiv.org/abs/1408.7049}{{\ttfamily arXiv:1408.7049 [nucl-th]}}.

\bibitem{Puglisi:2014pda}
A.~Puglisi, S.~Plumari, and V.~Greco, ``{Shear viscosity \ensuremath{\eta} to
  electric conductivity \ensuremath{\sigma}$_{el}$ ratio for the
  quark\textendash{}gluon plasma},''
  \href{http://dx.doi.org/10.1016/j.physletb.2015.10.070}{{\em Phys. Lett. B}
  {\bfseries 751} (2015) 326--330},
  \href{http://arxiv.org/abs/1407.2559}{{\ttfamily arXiv:1407.2559 [hep-ph]}}.

\bibitem{Puglisi:2014sha}
A.~Puglisi, S.~Plumari, and V.~Greco, ``{Electric Conductivity from the
  solution of the Relativistic Boltzmann Equation},''
  \href{http://dx.doi.org/10.1103/PhysRevD.90.114009}{{\em Phys. Rev. D}
  {\bfseries 90} (2014) 114009},
  \href{http://arxiv.org/abs/1408.7043}{{\ttfamily arXiv:1408.7043 [hep-ph]}}.

\bibitem{Cassing:2013iz}
W.~Cassing, O.~Linnyk, T.~Steinert, and V.~Ozvenchuk, ``{Electrical
  Conductivity of Hot QCD Matter},''
  \href{http://dx.doi.org/10.1103/PhysRevLett.110.182301}{{\em Phys. Rev.
  Lett.} {\bfseries 110} no.~18, (2013) 182301},
  \href{http://arxiv.org/abs/1302.0906}{{\ttfamily arXiv:1302.0906 [hep-ph]}}.

\bibitem{Steinert:2013fza}
T.~Steinert and W.~Cassing, ``{Electric and magnetic response of hot QCD
  matter},'' \href{http://dx.doi.org/10.1103/PhysRevC.89.035203}{{\em Phys.
  Rev. C} {\bfseries 89} no.~3, (2014) 035203},
  \href{http://arxiv.org/abs/1312.3189}{{\ttfamily arXiv:1312.3189 [hep-ph]}}.

\bibitem{Aarts:2014nba}
G.~Aarts, C.~Allton, A.~Amato, P.~Giudice, S.~Hands, and J.-I. Skullerud,
  ``{Electrical conductivity and charge diffusion in thermal QCD from the
  lattice},'' \href{http://dx.doi.org/10.1007/JHEP02(2015)186}{{\em JHEP}
  {\bfseries 02} (2015) 186}, \href{http://arxiv.org/abs/1412.6411}{{\ttfamily
  arXiv:1412.6411 [hep-lat]}}.

\bibitem{Aarts:2007wj}
G.~Aarts, C.~Allton, J.~Foley, S.~Hands, and S.~Kim, ``{Spectral functions at
  small energies and the electrical conductivity in hot, quenched lattice
  QCD},'' \href{http://dx.doi.org/10.1103/PhysRevLett.99.022002}{{\em Phys.
  Rev. Lett.} {\bfseries 99} (2007) 022002},
  \href{http://arxiv.org/abs/hep-lat/0703008}{{\ttfamily
  arXiv:hep-lat/0703008}}.

\bibitem{Amato:2013naa}
A.~Amato, G.~Aarts, C.~Allton, P.~Giudice, S.~Hands, and J.-I. Skullerud,
  ``{Electrical conductivity of the quark-gluon plasma across the deconfinement
  transition},'' \href{http://dx.doi.org/10.1103/PhysRevLett.111.172001}{{\em
  Phys. Rev. Lett.} {\bfseries 111} no.~17, (2013) 172001},
  \href{http://arxiv.org/abs/1307.6763}{{\ttfamily arXiv:1307.6763 [hep-lat]}}.

\bibitem{Gupta:2003zh}
S.~Gupta, ``{The Electrical conductivity and soft photon emissivity of the QCD
  plasma},'' \href{http://dx.doi.org/10.1016/j.physletb.2004.05.079}{{\em Phys.
  Lett. B} {\bfseries 597} (2004) 57--62},
  \href{http://arxiv.org/abs/hep-lat/0301006}{{\ttfamily
  arXiv:hep-lat/0301006}}.

\bibitem{Burnier:2012ts}
Y.~Burnier and M.~Laine, ``{Towards flavour diffusion coefficient and
  electrical conductivity without ultraviolet contamination},''
  \href{http://dx.doi.org/10.1140/epjc/s10052-012-1902-8}{{\em Eur. Phys. J. C}
  {\bfseries 72} (2012) 1902}, \href{http://arxiv.org/abs/1201.1994}{{\ttfamily
  arXiv:1201.1994 [hep-lat]}}.

\bibitem{Ding:2010ga}
H.~T. Ding, A.~Francis, O.~Kaczmarek, F.~Karsch, E.~Laermann, and W.~Soeldner,
  ``{Thermal dilepton rate and electrical conductivity: An analysis of vector
  current correlation functions in quenched lattice QCD},''
  \href{http://dx.doi.org/10.1103/PhysRevD.83.034504}{{\em Phys. Rev. D}
  {\bfseries 83} (2011) 034504},
  \href{http://arxiv.org/abs/1012.4963}{{\ttfamily arXiv:1012.4963 [hep-lat]}}.

\bibitem{Kaczmarek:2013dya}
O.~Kaczmarek and M.~M\"uller, ``{Temperature dependence of electrical
  conductivity and dilepton rates from hot quenched lattice QCD},''
  \href{http://dx.doi.org/10.22323/1.187.0175}{{\em PoS} {\bfseries
  LATTICE2013} (2014) 175}, \href{http://arxiv.org/abs/1312.5609}{{\ttfamily
  arXiv:1312.5609 [hep-lat]}}.

\bibitem{Marty:2013ita}
R.~Marty, E.~Bratkovskaya, W.~Cassing, J.~Aichelin, and H.~Berrehrah,
  ``{Transport coefficients from the Nambu-Jona-Lasinio model for $SU(3)_f$},''
  \href{http://dx.doi.org/10.1103/PhysRevC.88.045204}{{\em Phys. Rev. C}
  {\bfseries 88} (2013) 045204},
  \href{http://arxiv.org/abs/1305.7180}{{\ttfamily arXiv:1305.7180 [hep-ph]}}.

\bibitem{Aoki:2006we}
Y.~Aoki, G.~Endrodi, Z.~Fodor, S.~D. Katz, and K.~K. Szabo, ``{The Order of the
  quantum chromodynamics transition predicted by the standard model of particle
  physics},'' \href{http://dx.doi.org/10.1038/nature05120}{{\em Nature}
  {\bfseries 443} (2006) 675--678},
  \href{http://arxiv.org/abs/hep-lat/0611014}{{\ttfamily
  arXiv:hep-lat/0611014}}.

\bibitem{Asakawa:1989bq}
M.~Asakawa and K.~Yazaki, ``{Chiral Restoration at Finite Density and
  Temperature},'' \href{http://dx.doi.org/10.1016/0375-9474(89)90002-X}{{\em
  Nucl. Phys. A} {\bfseries 504} (1989) 668--684}.

\bibitem{Ejiri:2008xt}
S.~Ejiri, ``{Canonical partition function and finite density phase transition
  in lattice QCD},'' \href{http://dx.doi.org/10.1103/PhysRevD.78.074507}{{\em
  Phys. Rev. D} {\bfseries 78} (2008) 074507},
  \href{http://arxiv.org/abs/0804.3227}{{\ttfamily arXiv:0804.3227 [hep-lat]}}.

\bibitem{Stephanov:1999zu}
M.~A. Stephanov, K.~Rajagopal, and E.~V. Shuryak, ``{Event-by-event
  fluctuations in heavy ion collisions and the QCD critical point},''
  \href{http://dx.doi.org/10.1103/PhysRevD.60.114028}{{\em Phys. Rev. D}
  {\bfseries 60} (1999) 114028},
  \href{http://arxiv.org/abs/hep-ph/9903292}{{\ttfamily arXiv:hep-ph/9903292}}.

\bibitem{Hatta:2003wn}
Y.~Hatta and M.~Stephanov, ``{Proton number fluctuation as a signal of the QCD
  critical endpoint},''
  \href{http://dx.doi.org/10.1103/PhysRevLett.91.102003}{{\em Phys. Rev. Lett.}
  {\bfseries 91} (2003) 102003},
  \href{http://arxiv.org/abs/hep-ph/0302002}{{\ttfamily arXiv:hep-ph/0302002}}.
  [Erratum: Phys.Rev.Lett. 91, 129901 (2003)].

\bibitem{Asakawa:2000wh}
M.~Asakawa, U.~W. Heinz, and B.~Muller, ``{Fluctuation probes of quark
  deconfinement},'' \href{http://dx.doi.org/10.1103/PhysRevLett.85.2072}{{\em
  Phys. Rev. Lett.} {\bfseries 85} (2000) 2072--2075},
  \href{http://arxiv.org/abs/hep-ph/0003169}{{\ttfamily arXiv:hep-ph/0003169}}.

\bibitem{Jeon:1999gr}
S.~Jeon and V.~Koch, ``{Fluctuations of particle ratios and the abundance of
  hadronic resonances},''
  \href{http://dx.doi.org/10.1103/PhysRevLett.83.5435}{{\em Phys. Rev. Lett.}
  {\bfseries 83} (1999) 5435--5438},
  \href{http://arxiv.org/abs/nucl-th/9906074}{{\ttfamily
  arXiv:nucl-th/9906074}}.

\bibitem{Ejiri:2005wq}
S.~Ejiri, F.~Karsch, and K.~Redlich, ``{Hadronic fluctuations at the QCD phase
  transition},'' \href{http://dx.doi.org/10.1016/j.physletb.2005.11.083}{{\em
  Phys. Lett. B} {\bfseries 633} (2006) 275--282},
  \href{http://arxiv.org/abs/hep-ph/0509051}{{\ttfamily arXiv:hep-ph/0509051}}.

\bibitem{Kitazawa:2013bta}
M.~Kitazawa, M.~Asakawa, and H.~Ono, ``{Non-equilibrium time evolution of
  higher order cumulants of conserved charges and event-by-event analysis},''
  \href{http://dx.doi.org/10.1016/j.physletb.2013.12.008}{{\em Phys. Lett. B}
  {\bfseries 728} (2014) 386--392},
  \href{http://arxiv.org/abs/1307.2978}{{\ttfamily arXiv:1307.2978 [nucl-th]}}.

\bibitem{Skokov:2012ds}
V.~Skokov, B.~Friman, and K.~Redlich, ``{Volume Fluctuations and Higher Order
  Cumulants of the Net Baryon Number},''
  \href{http://dx.doi.org/10.1103/PhysRevC.88.034911}{{\em Phys. Rev. C}
  {\bfseries 88} (2013) 034911},
  \href{http://arxiv.org/abs/1205.4756}{{\ttfamily arXiv:1205.4756 [hep-ph]}}.

\bibitem{Pal:2020ucy}
S.~Pal, G.~Kadam, H.~Mishra, and A.~Bhattacharyya, ``{Effects of hadronic
  repulsive interactions on the fluctuations of conserved charges},''
  \href{http://dx.doi.org/10.1103/PhysRevD.103.054015}{{\em Phys. Rev. D}
  {\bfseries 103} no.~5, (2021) 054015},
  \href{http://arxiv.org/abs/2010.10761}{{\ttfamily arXiv:2010.10761
  [hep-ph]}}.

\bibitem{Asakawa:2015ybt}
M.~Asakawa and M.~Kitazawa, ``{Fluctuations of conserved charges in
  relativistic heavy ion collisions: An introduction},''
  \href{http://dx.doi.org/10.1016/j.ppnp.2016.04.002}{{\em Prog. Part. Nucl.
  Phys.} {\bfseries 90} (2016) 299--342},
  \href{http://arxiv.org/abs/1512.05038}{{\ttfamily arXiv:1512.05038
  [nucl-th]}}.

\bibitem{Jeon:2000wg}
S.~Jeon and V.~Koch, ``{Charged particle ratio fluctuation as a signal for
  QGP},'' \href{http://dx.doi.org/10.1103/PhysRevLett.85.2076}{{\em Phys. Rev.
  Lett.} {\bfseries 85} (2000) 2076--2079},
  \href{http://arxiv.org/abs/hep-ph/0003168}{{\ttfamily arXiv:hep-ph/0003168}}.

\bibitem{Shuryak:2000pd}
E.~V. Shuryak and M.~A. Stephanov, ``{When can long range charge fluctuations
  serve as a QGP signal?},''
  \href{http://dx.doi.org/10.1103/PhysRevC.63.064903}{{\em Phys. Rev. C}
  {\bfseries 63} (2001) 064903},
  \href{http://arxiv.org/abs/hep-ph/0010100}{{\ttfamily arXiv:hep-ph/0010100}}.

\bibitem{Pratt:2019pnd}
S.~Pratt and C.~Plumberg, ``{Determining the Diffusivity for Light Quarks from
  Experiment},'' \href{http://dx.doi.org/10.1103/PhysRevC.102.044909}{{\em
  Phys. Rev. C} {\bfseries 102} no.~4, (2020) 044909},
  \href{http://arxiv.org/abs/1904.11459}{{\ttfamily arXiv:1904.11459
  [nucl-th]}}.

\bibitem{Fotakis:2019nbq}
J.~A. Fotakis, M.~Greif, C.~Greiner, G.~S. Denicol, and H.~Niemi, ``{Diffusion
  processes involving multiple conserved charges: A study from kinetic theory
  and implications to the fluid-dynamical modeling of heavy ion collisions},''
  \href{http://dx.doi.org/10.1103/PhysRevD.101.076007}{{\em Phys. Rev. D}
  {\bfseries 101} no.~7, (2020) 076007},
  \href{http://arxiv.org/abs/1912.09103}{{\ttfamily arXiv:1912.09103
  [hep-ph]}}.

\bibitem{Monnai:2012jc}
A.~Monnai, ``{Dissipative Hydrodynamic Effects on Baryon Stopping},''
  \href{http://dx.doi.org/10.1103/PhysRevC.86.014908}{{\em Phys. Rev. C}
  {\bfseries 86} (2012) 014908},
  \href{http://arxiv.org/abs/1204.4713}{{\ttfamily arXiv:1204.4713 [nucl-th]}}.

\bibitem{STAR:2010vob}
{\bfseries STAR} Collaboration, M.~M. Aggarwal {\em et~al.}, ``{An Experimental
  Exploration of the QCD Phase Diagram: The Search for the Critical Point and
  the Onset of De-confinement},''
  \href{http://arxiv.org/abs/1007.2613}{{\ttfamily arXiv:1007.2613 [nucl-ex]}}.

\bibitem{Mohanty:2011nm}
{\bfseries STAR} Collaboration, B.~Mohanty, ``{STAR experiment results from the
  beam energy scan program at RHIC},''
  \href{http://dx.doi.org/10.1088/0954-3899/38/12/124023}{{\em J. Phys. G}
  {\bfseries 38} (2011) 124023},
  \href{http://arxiv.org/abs/1106.5902}{{\ttfamily arXiv:1106.5902 [nucl-ex]}}.

\bibitem{Mitchell:2012mx}
{\bfseries PHENIX} Collaboration, J.~T. Mitchell, ``{The RHIC Beam Energy Scan
  Program: Results from the PHENIX Experiment},''
  \href{http://dx.doi.org/10.1016/j.nuclphysa.2013.02.161}{{\em Nucl. Phys. A}
  {\bfseries 904-905} (2013) 903c--906c},
  \href{http://arxiv.org/abs/1211.6139}{{\ttfamily arXiv:1211.6139 [nucl-ex]}}.

\bibitem{Odyniec:2013kna}
G.~Odyniec, ``{The RHIC Beam Energy Scan program in STAR and what's next ...}''
  \href{http://dx.doi.org/10.1088/1742-6596/455/1/012037}{{\em J. Phys. Conf.
  Ser.} {\bfseries 455} (2013) 012037}.

\bibitem{STAR:2017sal}
{\bfseries STAR} Collaboration, L.~Adamczyk {\em et~al.}, ``{Bulk Properties of
  the Medium Produced in Relativistic Heavy-Ion Collisions from the Beam Energy
  Scan Program},'' \href{http://dx.doi.org/10.1103/PhysRevC.96.044904}{{\em
  Phys. Rev. C} {\bfseries 96} no.~4, (2017) 044904},
  \href{http://arxiv.org/abs/1701.07065}{{\ttfamily arXiv:1701.07065
  [nucl-ex]}}.

\bibitem{Friman:2011zz}
B.~Friman, C.~Hohne, J.~Knoll, S.~Leupold, J.~Randrup, R.~Rapp, and P.~Senger,
  eds., \href{http://dx.doi.org/10.1007/978-3-642-13293-3}{{\em {The CBM
  physics book: Compressed baryonic matter in laboratory experiments}}},
  vol.~814.
\newblock 2011.

\bibitem{Greif:2017byw}
M.~Greif, J.~A. Fotakis, G.~S. Denicol, and C.~Greiner, ``{Diffusion of
  conserved charges in relativistic heavy ion collisions},''
  \href{http://dx.doi.org/10.1103/PhysRevLett.120.242301}{{\em Phys. Rev.
  Lett.} {\bfseries 120} no.~24, (2018) 242301},
  \href{http://arxiv.org/abs/1711.08680}{{\ttfamily arXiv:1711.08680
  [hep-ph]}}.

\bibitem{Braun-Munzinger:2003pwq}
P.~Braun-Munzinger, K.~Redlich, and J.~Stachel, ``{Particle production in heavy
  ion collisions},'' \href{http://arxiv.org/abs/nucl-th/0304013}{{\ttfamily
  arXiv:nucl-th/0304013}}.

\bibitem{Andronic:2005yp}
A.~Andronic, P.~Braun-Munzinger, and J.~Stachel, ``{Hadron production in
  central nucleus-nucleus collisions at chemical freeze-out},''
  \href{http://dx.doi.org/10.1016/j.nuclphysa.2006.03.012}{{\em Nucl. Phys. A}
  {\bfseries 772} (2006) 167--199},
  \href{http://arxiv.org/abs/nucl-th/0511071}{{\ttfamily
  arXiv:nucl-th/0511071}}.

\bibitem{PhysRev.187.345}
R.~Dashen, S.-k. Ma, and H.~J. Bernstein, ``$s$-matrix formulation of
  statistical mechanics,''
  \href{http://dx.doi.org/10.1103/PhysRev.187.345}{{\em Phys. Rev.} {\bfseries
  187} (Nov, 1969) 345--370}.

\bibitem{PhysRevD.10.694}
R.~F. Dashen and R.~Rajaraman, ``Narrow resonances in statistical mechanics,''
  \href{http://dx.doi.org/10.1103/PhysRevD.10.694}{{\em Phys. Rev. D}
  {\bfseries 10} (Jul, 1974) 694--708}.

\bibitem{Braun-Munzinger:2001hwo}
P.~Braun-Munzinger, D.~Magestro, K.~Redlich, and J.~Stachel, ``{Hadron
  production in Au - Au collisions at RHIC},''
  \href{http://dx.doi.org/10.1016/S0370-2693(01)01069-3}{{\em Phys. Lett. B}
  {\bfseries 518} (2001) 41--46},
  \href{http://arxiv.org/abs/hep-ph/0105229}{{\ttfamily arXiv:hep-ph/0105229}}.

\bibitem{Cleymans:1999st}
J.~Cleymans and K.~Redlich, ``{Chemical and thermal freezeout parameters from
  1-A/GeV to 200-A/GeV},''
  \href{http://dx.doi.org/10.1103/PhysRevC.60.054908}{{\em Phys. Rev. C}
  {\bfseries 60} (1999) 054908},
  \href{http://arxiv.org/abs/nucl-th/9903063}{{\ttfamily
  arXiv:nucl-th/9903063}}.

\bibitem{Becattini:2000jw}
F.~Becattini, J.~Cleymans, A.~Keranen, E.~Suhonen, and K.~Redlich, ``{Features
  of particle multiplicities and strangeness production in central heavy ion
  collisions between 1.7A-GeV/c and 158A-GeV/c},''
  \href{http://dx.doi.org/10.1103/PhysRevC.64.024901}{{\em Phys. Rev. C}
  {\bfseries 64} (2001) 024901},
  \href{http://arxiv.org/abs/hep-ph/0002267}{{\ttfamily arXiv:hep-ph/0002267}}.

\bibitem{Cleymans:2004pp}
J.~Cleymans, B.~Kampfer, M.~Kaneta, S.~Wheaton, and N.~Xu, ``{Centrality
  dependence of thermal parameters deduced from hadron multiplicities in Au +
  Au collisions at s(NN)**(1/2) = 130-GeV},''
  \href{http://dx.doi.org/10.1103/PhysRevC.71.054901}{{\em Phys. Rev. C}
  {\bfseries 71} (2005) 054901},
  \href{http://arxiv.org/abs/hep-ph/0409071}{{\ttfamily arXiv:hep-ph/0409071}}.

\bibitem{Andronic:2008gu}
A.~Andronic, P.~Braun-Munzinger, and J.~Stachel, ``{Thermal hadron production
  in relativistic nuclear collisions: The Hadron mass spectrum, the horn, and
  the QCD phase transition},''
  \href{http://dx.doi.org/10.1016/j.physletb.2009.06.021}{{\em Phys. Lett. B}
  {\bfseries 673} (2009) 142--145},
  \href{http://arxiv.org/abs/0812.1186}{{\ttfamily arXiv:0812.1186 [nucl-th]}}.
  [Erratum: Phys.Lett.B 678, 516 (2009)].

\bibitem{Karsch:2003zq}
F.~Karsch, K.~Redlich, and A.~Tawfik, ``{Thermodynamics at nonzero baryon
  number density: A Comparison of lattice and hadron resonance gas model
  calculations},'' \href{http://dx.doi.org/10.1016/j.physletb.2003.08.001}{{\em
  Phys. Lett. B} {\bfseries 571} (2003) 67--74},
  \href{http://arxiv.org/abs/hep-ph/0306208}{{\ttfamily arXiv:hep-ph/0306208}}.

\bibitem{Braun-Munzinger:2015hba}
P.~Braun-Munzinger, V.~Koch, T.~Sch\"afer, and J.~Stachel, ``{Properties of hot
  and dense matter from relativistic heavy ion collisions},''
  \href{http://dx.doi.org/10.1016/j.physrep.2015.12.003}{{\em Phys. Rept.}
  {\bfseries 621} (2016) 76--126},
  \href{http://arxiv.org/abs/1510.00442}{{\ttfamily arXiv:1510.00442
  [nucl-th]}}.

\bibitem{Nahrgang:2014fza}
M.~Nahrgang, M.~Bluhm, P.~Alba, R.~Bellwied, and C.~Ratti, ``{Impact of
  resonance regeneration and decay on the net-proton fluctuations in a hadron
  resonance gas},''
  \href{http://dx.doi.org/10.1140/epjc/s10052-015-3775-0}{{\em Eur. Phys. J. C}
  {\bfseries 75} no.~12, (2015) 573},
  \href{http://arxiv.org/abs/1402.1238}{{\ttfamily arXiv:1402.1238 [hep-ph]}}.

\bibitem{Bhattacharyya:2013oya}
A.~Bhattacharyya, S.~Das, S.~K. Ghosh, R.~Ray, and S.~Samanta, ``{Fluctuations
  and correlations of conserved charges in an excluded volume hadron resonance
  gas model},'' \href{http://dx.doi.org/10.1103/PhysRevC.90.034909}{{\em Phys.
  Rev. C} {\bfseries 90} no.~3, (2014) 034909},
  \href{http://arxiv.org/abs/1310.2793}{{\ttfamily arXiv:1310.2793 [hep-ph]}}.

\bibitem{Garg:2013ata}
P.~Garg, D.~K. Mishra, P.~K. Netrakanti, B.~Mohanty, A.~K. Mohanty, B.~K.
  Singh, and N.~Xu, ``{Conserved number fluctuations in a hadron resonance gas
  model},'' \href{http://dx.doi.org/10.1016/j.physletb.2013.09.019}{{\em Phys.
  Lett. B} {\bfseries 726} (2013) 691--696},
  \href{http://arxiv.org/abs/1304.7133}{{\ttfamily arXiv:1304.7133 [nucl-ex]}}.

\bibitem{PhysRevD.86.034509}
{\bfseries HotQCD Collaboration} Collaboration, A.~Bazavov, T.~Bhattacharya,
  C.~E. DeTar, H.-T. Ding, S.~Gottlieb, R.~Gupta, P.~Hegde, U.~M. Heller,
  F.~Karsch, E.~Laermann, L.~Levkova, S.~Mukherjee, P.~Petreczky, C.~Schmidt,
  R.~A. Soltz, W.~Soeldner, R.~Sugar, and P.~M. Vranas, ``Fluctuations and
  correlations of net baryon number, electric charge, and strangeness: A
  comparison of lattice qcd results with the hadron resonance gas model,''
  \href{http://dx.doi.org/10.1103/PhysRevD.86.034509}{{\em Phys. Rev. D}
  {\bfseries 86} (Aug, 2012) 034509}.

\bibitem{Begun:2006jf}
V.~V. Begun, M.~I. Gorenstein, M.~Hauer, V.~P. Konchakovski, and O.~S. Zozulya,
  ``{Multiplicity Fluctuations in Hadron-Resonance Gas},''
  \href{http://dx.doi.org/10.1103/PhysRevC.74.044903}{{\em Phys. Rev. C}
  {\bfseries 74} (2006) 044903},
  \href{http://arxiv.org/abs/nucl-th/0606036}{{\ttfamily
  arXiv:nucl-th/0606036}}.

\bibitem{Prakash:1993bt}
M.~Prakash, M.~Prakash, R.~Venugopalan, and G.~Welke, ``{Nonequilibrium
  properties of hadronic mixtures},''
  \href{http://dx.doi.org/10.1016/0370-1573(93)90092-R}{{\em Phys. Rept.}
  {\bfseries 227} (1993) 321--366}.

\bibitem{Wiranata:2012br}
A.~Wiranata and M.~Prakash, ``{Shear Viscosities from the Chapman-Enskog and
  the Relaxation Time Approaches},''
  \href{http://dx.doi.org/10.1103/PhysRevC.85.054908}{{\em Phys. Rev. C}
  {\bfseries 85} (2012) 054908},
  \href{http://arxiv.org/abs/1203.0281}{{\ttfamily arXiv:1203.0281 [nucl-th]}}.

\bibitem{PhysRevC.83.014906}
P.~Chakraborty and J.~I. Kapusta, ``Quasiparticle theory of shear and bulk
  viscosities of hadronic matter,''
  \href{http://dx.doi.org/10.1103/PhysRevC.83.014906}{{\em Phys. Rev. C}
  {\bfseries 83} (Jan, 2011) 014906}.

\bibitem{Khvorostukhin:2010aj}
A.~S. Khvorostukhin, V.~D. Toneev, and D.~N. Voskresensky, ``{Viscosity
  Coefficients for Hadron and Quark-Gluon Phases},''
  \href{http://dx.doi.org/10.1016/j.nuclphysa.2010.05.058}{{\em Nucl. Phys. A}
  {\bfseries 845} (2010) 106--146},
  \href{http://arxiv.org/abs/1003.3531}{{\ttfamily arXiv:1003.3531 [nucl-th]}}.

\bibitem{PhysRevC.86.054902}
S.~Plumari, A.~Puglisi, F.~Scardina, and V.~Greco, ``Shear viscosity of a
  strongly interacting system: Green-kubo correlator versus chapman-enskog and
  relaxation-time approximations,''
  \href{http://dx.doi.org/10.1103/PhysRevC.86.054902}{{\em Phys. Rev. C}
  {\bfseries 86} (Nov, 2012) 054902}.
  \url{https://link.aps.org/doi/10.1103/PhysRevC.86.054902}.

\bibitem{PhysRevC.77.024911}
M.~I. Gorenstein, M.~Hauer, and O.~N. Moroz, ``Viscosity in the excluded volume
  hadron gas model,'' \href{http://dx.doi.org/10.1103/PhysRevC.77.024911}{{\em
  Phys. Rev. C} {\bfseries 77} (Feb, 2008) 024911}.
  \url{https://link.aps.org/doi/10.1103/PhysRevC.77.024911}.

\bibitem{PhysRevC.86.024913}
J.~Noronha-Hostler, J.~Noronha, and C.~Greiner, ``Hadron mass spectrum and the
  shear viscosity to entropy density ratio of hot hadronic matter,''
  \href{http://dx.doi.org/10.1103/PhysRevC.86.024913}{{\em Phys. Rev. C}
  {\bfseries 86} (Aug, 2012) 024913}.
  \url{https://link.aps.org/doi/10.1103/PhysRevC.86.024913}.

\bibitem{PhysRevC.85.014908}
S.~K. Tiwari, P.~K. Srivastava, and C.~P. Singh, ``Description of hot and dense
  hadron-gas properties in a new excluded-volume model,''
  \href{http://dx.doi.org/10.1103/PhysRevC.85.014908}{{\em Phys. Rev. C}
  {\bfseries 85} (Jan, 2012) 014908}.
  \url{https://link.aps.org/doi/10.1103/PhysRevC.85.014908}.

\bibitem{PhysRevC.88.068201}
S.~Ghosh, A.~Lahiri, S.~Majumder, R.~Ray, and S.~K. Ghosh, ``Shear viscosity
  due to landau damping from the quark-pion interaction,''
  \href{http://dx.doi.org/10.1103/PhysRevC.88.068201}{{\em Phys. Rev. C}
  {\bfseries 88} (Dec, 2013) 068201}.
  \url{https://link.aps.org/doi/10.1103/PhysRevC.88.068201}.

\bibitem{PhysRevC.89.045201}
S.~Ghosh, G.~a. Krein, and S.~Sarkar, ``Shear viscosity of a pion gas resulting
  from $\ensuremath{\rho}\ensuremath{\pi}\ensuremath{\pi}$ and
  $\ensuremath{\sigma}\ensuremath{\pi}\ensuremath{\pi}$ interactions,''
  \href{http://dx.doi.org/10.1103/PhysRevC.89.045201}{{\em Phys. Rev. C}
  {\bfseries 89} (Apr, 2014) 045201}.
  \url{https://link.aps.org/doi/10.1103/PhysRevC.89.045201}.

\bibitem{Wiranata:2014kva}
A.~Wiranata, V.~Koch, M.~Prakash, and X.~N. Wang, ``{Shear viscosity of a
  multi-component hadronic system},''
  \href{http://dx.doi.org/10.1088/1742-6596/509/1/012049}{{\em J. Phys. Conf.
  Ser.} {\bfseries 509} (2014) 012049}.

\bibitem{Wiranata:2012vv}
A.~Wiranata, M.~Prakash, and P.~Chakraborty, ``{Comparison of Viscosities from
  the Chapman-Enskog and Relaxation Time Methods},''
  \href{http://dx.doi.org/10.2478/s11534-012-0082-3}{{\em Central Eur. J.
  Phys.} {\bfseries 10} (2012) 1349--1351},
  \href{http://arxiv.org/abs/1201.3104}{{\ttfamily arXiv:1201.3104 [nucl-th]}}.

\bibitem{Tawfik:2010mb}
A.~Tawfik and M.~Wahba, ``{Bulk and Shear Viscosity in Hagedorn Fluid},''
  \href{http://dx.doi.org/10.1002/andp.201000056}{{\em Annalen Phys.}
  {\bfseries 522} (2010) 849--856},
  \href{http://arxiv.org/abs/1005.3946}{{\ttfamily arXiv:1005.3946 [hep-ph]}}.

\bibitem{PhysRevLett.103.172302}
J.~Noronha-Hostler, J.~Noronha, and C.~Greiner, ``Transport coefficients of
  hadronic matter near ${T}_{c}$,''
  \href{http://dx.doi.org/10.1103/PhysRevLett.103.172302}{{\em Phys. Rev.
  Lett.} {\bfseries 103} (Oct, 2009) 172302}.
  \url{https://link.aps.org/doi/10.1103/PhysRevLett.103.172302}.

\bibitem{Kadam:2014cua}
G.~P. Kadam and H.~Mishra, ``{Bulk and shear viscosities of hot and dense
  hadron gas},'' \href{http://dx.doi.org/10.1016/j.nuclphysa.2014.12.004}{{\em
  Nucl. Phys. A} {\bfseries 934} (2014) 133--147},
  \href{http://arxiv.org/abs/1408.6329}{{\ttfamily arXiv:1408.6329 [hep-ph]}}.

\bibitem{Kadam:2014xka}
G.~Kadam, ``{Transport properties of hadronic matter in magnetic field},''
  \href{http://dx.doi.org/10.1142/S0217732315500315}{{\em Mod. Phys. Lett. A}
  {\bfseries 30} no.~10, (2015) 1550031},
  \href{http://arxiv.org/abs/1412.5303}{{\ttfamily arXiv:1412.5303 [hep-ph]}}.

\bibitem{Ghosh:2014yea}
S.~Ghosh, ``{A real-time thermal field theoretical analysis of Kubo-type shear
  viscosity: Numerical understanding with simple examples},''
  \href{http://dx.doi.org/10.1142/S0217751X14500547}{{\em Int. J. Mod. Phys. A}
  {\bfseries 29} (2014) 1450054},
  \href{http://arxiv.org/abs/1404.4788}{{\ttfamily arXiv:1404.4788 [nucl-th]}}.

\bibitem{Demir:2014kda}
N.~Demir and A.~Wiranata, ``{Hadronic Shear Viscosity: A Comparison between the
  Green-Kubo and Chapmann Enskog Methods},''
  \href{http://dx.doi.org/10.1088/1742-6596/535/1/012018}{{\em J. Phys. Conf.
  Ser.} {\bfseries 535} (2014) 012018}.

\bibitem{PhysRevC.90.025202}
S.~Ghosh, ``Nucleon thermal width owing to pion-baryon loops and its
  contributions to shear viscosity,''
  \href{http://dx.doi.org/10.1103/PhysRevC.90.025202}{{\em Phys. Rev. C}
  {\bfseries 90} (Aug, 2014) 025202}.
  \url{https://link.aps.org/doi/10.1103/PhysRevC.90.025202}.

\bibitem{PhysRevC.97.055204}
J.-B. Rose, J.~M. Torres-Rincon, A.~Sch\"afer, D.~R. Oliinychenko, and
  H.~Petersen, ``Shear viscosity of a hadron gas and influence of resonance
  lifetimes on relaxation time,''
  \href{http://dx.doi.org/10.1103/PhysRevC.97.055204}{{\em Phys. Rev. C}
  {\bfseries 97} (May, 2018) 055204}.
  \url{https://link.aps.org/doi/10.1103/PhysRevC.97.055204}.

\bibitem{PhysRevC.84.054911}
C.~Wesp, A.~El, F.~Reining, Z.~Xu, I.~Bouras, and C.~Greiner, ``Calculation of
  shear viscosity using green-kubo relations within a parton cascade,''
  \href{http://dx.doi.org/10.1103/PhysRevC.84.054911}{{\em Phys. Rev. C}
  {\bfseries 84} (Nov, 2011) 054911}.
  \url{https://link.aps.org/doi/10.1103/PhysRevC.84.054911}.

\bibitem{PhysRevD.90.094014}
M.~Greif, I.~Bouras, C.~Greiner, and Z.~Xu, ``Electric conductivity of the
  quark-gluon plasma investigated using a perturbative qcd based parton
  cascade,'' \href{http://dx.doi.org/10.1103/PhysRevD.90.094014}{{\em Phys.
  Rev. D} {\bfseries 90} (Nov, 2014) 094014}.
  \url{https://link.aps.org/doi/10.1103/PhysRevD.90.094014}.

\bibitem{Bass:1998ca}
S.~A. Bass {\em et~al.}, ``{Microscopic models for ultrarelativistic heavy ion
  collisions},'' \href{http://dx.doi.org/10.1016/S0146-6410(98)00058-1}{{\em
  Prog. Part. Nucl. Phys.} {\bfseries 41} (1998) 255--369},
  \href{http://arxiv.org/abs/nucl-th/9803035}{{\ttfamily
  arXiv:nucl-th/9803035}}.

\bibitem{PhysRevC.92.035203}
G.~P. Kadam and H.~Mishra, ``Dissipative properties of hot and dense hadronic
  matter in an excluded-volume hadron resonance gas model,''
  \href{http://dx.doi.org/10.1103/PhysRevC.92.035203}{{\em Phys. Rev. C}
  {\bfseries 92} (Sep, 2015) 035203}.
  \url{https://link.aps.org/doi/10.1103/PhysRevC.92.035203}.

\bibitem{PhysRevD.99.014015}
J.~R. Bhatt, A.~Das, and H.~Mishra, ``Thermoelectric effect and seebeck
  coefficient for hot and dense hadronic matter,''
  \href{http://dx.doi.org/10.1103/PhysRevD.99.014015}{{\em Phys. Rev. D}
  {\bfseries 99} (Jan, 2019) 014015}.
  \url{https://link.aps.org/doi/10.1103/PhysRevD.99.014015}.

\bibitem{Das:2021qii}
A.~Das and H.~Mishra, ``{Thermoelectric transport coefficients of hot and dense
  QCD matter},'' \href{http://dx.doi.org/10.1140/epjs/s11734-021-00022-2}{{\em
  Eur. Phys. J. ST} {\bfseries 230} no.~3, (2021) 607--634}.

\bibitem{Mohapatra:2019mcl}
R.~K. Mohapatra, H.~Mishra, S.~Dash, and B.~K. Nandi, ``{Transport coefficients
  of hadronic matter in a van der Walls hadron resonance gas model},''
  \href{http://arxiv.org/abs/1901.07238}{{\ttfamily arXiv:1901.07238
  [hep-ph]}}.

\bibitem{Fotakis:2021diq}
J.~A. Fotakis, O.~Soloveva, C.~Greiner, O.~Kaczmarek, and E.~Bratkovskaya,
  ``{Diffusion coefficient matrix of the strongly interacting quark-gluon
  plasma},'' \href{http://arxiv.org/abs/2102.08140}{{\ttfamily arXiv:2102.08140
  [hep-ph]}}.

\bibitem{Chakraborty:2010fr}
P.~Chakraborty and J.~I. Kapusta, ``{Quasi-Particle Theory of Shear and Bulk
  Viscosities of Hadronic Matter},''
  \href{http://dx.doi.org/10.1103/PhysRevC.83.014906}{{\em Phys. Rev. C}
  {\bfseries 83} (2011) 014906},
  \href{http://arxiv.org/abs/1006.0257}{{\ttfamily arXiv:1006.0257 [nucl-th]}}.

\bibitem{Albright:2015fpa}
M.~Albright and J.~I. Kapusta, ``{Quasiparticle Theory of Transport
  Coefficients for Hadronic Matter at Finite Temperature and Baryon Density},''
  \href{http://dx.doi.org/10.1103/PhysRevC.93.014903}{{\em Phys. Rev. C}
  {\bfseries 93} no.~1, (2016) 014903},
  \href{http://arxiv.org/abs/1508.02696}{{\ttfamily arXiv:1508.02696
  [nucl-th]}}.

\bibitem{Albright:2015edp}
M.~Albright, {\em {Thermodynamics of Hot Hadronic Gases at Finite Baryon
  Densities}}.
\newblock PhD thesis, Minnesota U., 2015.

\bibitem{Deb:2016myz}
P.~Deb, G.~P. Kadam, and H.~Mishra, ``{Estimating transport coefficients in hot
  and dense quark matter},''
  \href{http://dx.doi.org/10.1103/PhysRevD.94.094002}{{\em Phys. Rev. D}
  {\bfseries 94} no.~9, (2016) 094002},
  \href{http://arxiv.org/abs/1603.01952}{{\ttfamily arXiv:1603.01952
  [hep-ph]}}.

\bibitem{Ollitrault:2007du}
J.-Y. Ollitrault, ``{Relativistic hydrodynamics for heavy-ion collisions},''
  \href{http://dx.doi.org/10.1088/0143-0807/29/2/010}{{\em Eur. J. Phys.}
  {\bfseries 29} (2008) 275--302},
  \href{http://arxiv.org/abs/0708.2433}{{\ttfamily arXiv:0708.2433 [nucl-th]}}.

\bibitem{PhysRev.37.405}
L.~Onsager, ``Reciprocal relations in irreversible processes. i.''
  \href{http://dx.doi.org/10.1103/PhysRev.37.405}{{\em Phys. Rev.} {\bfseries
  37} (Feb, 1931) 405--426}.

\bibitem{PhysRev.38.2265}
L.~Onsager, ``Reciprocal relations in irreversible processes. ii.''
  \href{http://dx.doi.org/10.1103/PhysRev.38.2265}{{\em Phys. Rev.} {\bfseries
  38} (Dec, 1931) 2265--2279}.

\bibitem{PhysRev.94.511}
P.~L. Bhatnagar, E.~P. Gross, and M.~Krook, ``A model for collision processes
  in gases. i. small amplitude processes in charged and neutral one-component
  systems,'' \href{http://dx.doi.org/10.1103/PhysRev.94.511}{{\em Phys. Rev.}
  {\bfseries 94} (May, 1954) 511--525}.
  \url{https://link.aps.org/doi/10.1103/PhysRev.94.511}.

\bibitem{Rocha:2021zcw}
G.~S. Rocha, G.~S. Denicol, and J.~Noronha, ``{Novel Relaxation Time
  Approximation to the Relativistic Boltzmann Equation},''
  \href{http://arxiv.org/abs/2103.07489}{{\ttfamily arXiv:2103.07489
  [nucl-th]}}.

\bibitem{PhysRevC.88.024902}
V.~V. Begun, M.~Ga\ifmmode~\acute{z}\else \'{z}\fi{}dzicki, and M.~I.
  Gorenstein, ``Hadron-resonance gas at freeze-out: Reminder on the importance
  of repulsive interactions,''
  \href{http://dx.doi.org/10.1103/PhysRevC.88.024902}{{\em Phys. Rev. C}
  {\bfseries 88} (Aug, 2013) 024902}.

\bibitem{Andronic:2012ut}
A.~Andronic, P.~Braun-Munzinger, J.~Stachel, and M.~Winn, ``{Interacting hadron
  resonance gas meets lattice QCD},''
  \href{http://dx.doi.org/10.1016/j.physletb.2012.10.001}{{\em Phys. Lett. B}
  {\bfseries 718} (2012) 80--85},
  \href{http://arxiv.org/abs/1201.0693}{{\ttfamily arXiv:1201.0693 [nucl-th]}}.

\bibitem{Rischke:1991ke}
D.~H. Rischke, M.~I. Gorenstein, H.~Stoecker, and W.~Greiner, ``{Excluded
  volume effect for the nuclear matter equation of state},''
  \href{http://dx.doi.org/10.1007/BF01548574}{{\em Z. Phys. C} {\bfseries 51}
  (1991) 485--490}.

\bibitem{Hama:2004rr}
Y.~Hama, T.~Kodama, and O.~Socolowski, Jr., ``{Topics on hydrodynamic model of
  nucleus-nucleus collisions},''
  \href{http://dx.doi.org/10.1590/S0103-97332005000100003}{{\em Braz. J. Phys.}
  {\bfseries 35} (2005) 24--51},
  \href{http://arxiv.org/abs/hep-ph/0407264}{{\ttfamily arXiv:hep-ph/0407264}}.

\bibitem{PhysRevC.82.044904}
K.~Werner, I.~Karpenko, T.~Pierog, M.~Bleicher, and K.~Mikhailov,
  ``Event-by-event simulation of the three-dimensional hydrodynamic evolution
  from flux tube initial conditions in ultrarelativistic heavy ion
  collisions,'' \href{http://dx.doi.org/10.1103/PhysRevC.82.044904}{{\em Phys.
  Rev. C} {\bfseries 82} (Oct, 2010) 044904}.
  \url{https://link.aps.org/doi/10.1103/PhysRevC.82.044904}.

\bibitem{Kadam:2015xsa}
G.~P. Kadam and H.~Mishra, ``{Dissipative properties of hot and dense hadronic
  matter in an excluded-volume hadron resonance gas model},''
  \href{http://dx.doi.org/10.1103/PhysRevC.92.035203}{{\em Phys. Rev. C}
  {\bfseries 92} no.~3, (2015) 035203},
  \href{http://arxiv.org/abs/1506.04613}{{\ttfamily arXiv:1506.04613
  [hep-ph]}}.

\bibitem{Braun-Munzinger:1994ewq}
P.~Braun-Munzinger, J.~Stachel, J.~P. Wessels, and N.~Xu, ``{Thermal
  equilibration and expansion in nucleus-nucleus collisions at the AGS},''
  \href{http://dx.doi.org/10.1016/0370-2693(94)01534-J}{{\em Phys. Lett. B}
  {\bfseries 344} (1995) 43--48},
  \href{http://arxiv.org/abs/nucl-th/9410026}{{\ttfamily
  arXiv:nucl-th/9410026}}.

\bibitem{Hagedorn:1980kb}
R.~Hagedorn and J.~Rafelski, ``{Hot Hadronic Matter and Nuclear Collisions},''
  \href{http://dx.doi.org/10.1016/0370-2693(80)90566-3}{{\em Phys. Lett. B}
  {\bfseries 97} (1980) 136}.

\bibitem{Cleymans:1992jz}
J.~Cleymans, M.~I. Gorenstein, J.~Stalnacke, and E.~Suhonen, ``{Excluded volume
  effect and the quark - hadron phase transition},''
  \href{http://dx.doi.org/10.1088/0031-8949/48/3/004}{{\em Phys. Scripta}
  {\bfseries 48} (1993) 277--280}.

\bibitem{PhysRevC.56.2210}
G.~D. Yen, M.~I. Gorenstein, W.~Greiner, and S.~N. Yang, ``Excluded volume
  hadron gas model for particle number ratios in $\mathit{A}+\mathit{A}$
  collisions,'' \href{http://dx.doi.org/10.1103/PhysRevC.56.2210}{{\em Phys.
  Rev. C} {\bfseries 56} (Oct, 1997) 2210--2218}.
  \url{https://link.aps.org/doi/10.1103/PhysRevC.56.2210}.

\bibitem{ParticleDataGroup:2008zun}
{\bfseries Particle Data Group} Collaboration, C.~Amsler {\em et~al.},
  ``{Review of Particle Physics},''
  \href{http://dx.doi.org/10.1016/j.physletb.2008.07.018}{{\em Phys. Lett. B}
  {\bfseries 667} (2008) 1--1340}.

\bibitem{PhysRevC.90.024915}
M.~Albright, J.~Kapusta, and C.~Young, ``Matching excluded-volume
  hadron-resonance gas models and perturbative qcd to lattice calculations,''
  \href{http://dx.doi.org/10.1103/PhysRevC.90.024915}{{\em Phys. Rev. C}
  {\bfseries 90} (Aug, 2014) 024915}.

\bibitem{Braun-Munzinger:1999hun}
P.~Braun-Munzinger, I.~Heppe, and J.~Stachel, ``{Chemical equilibration in Pb +
  Pb collisions at the SPS},''
  \href{http://dx.doi.org/10.1016/S0370-2693(99)01076-X}{{\em Phys. Lett. B}
  {\bfseries 465} (1999) 15--20},
  \href{http://arxiv.org/abs/nucl-th/9903010}{{\ttfamily
  arXiv:nucl-th/9903010}}.

\bibitem{Das:2019pqd}
A.~Das, H.~Mishra, and R.~K. Mohapatra, ``{Transport coefficients of hot and
  dense hadron gas in a magnetic field: a relaxation time approach},''
  \href{http://dx.doi.org/10.1103/PhysRevD.100.114004}{{\em Phys. Rev. D}
  {\bfseries 100} no.~11, (2019) 114004},
  \href{http://arxiv.org/abs/1909.06202}{{\ttfamily arXiv:1909.06202
  [hep-ph]}}.

\bibitem{Das:2019wjg}
A.~Das, H.~Mishra, and R.~K. Mohapatra, ``{Electrical conductivity and Hall
  conductivity of a hot and dense hadron gas in a magnetic field: A relaxation
  time approach},'' \href{http://dx.doi.org/10.1103/PhysRevD.99.094031}{{\em
  Phys. Rev. D} {\bfseries 99} no.~9, (2019) 094031},
  \href{http://arxiv.org/abs/1903.03938}{{\ttfamily arXiv:1903.03938
  [hep-ph]}}.

\bibitem{Florkowski:2014sfa}
W.~Florkowski, E.~Maksymiuk, R.~Ryblewski, and M.~Strickland, ``{Exact solution
  of the (0+1)-dimensional Boltzmann equation for a massive gas},''
  \href{http://dx.doi.org/10.1103/PhysRevC.89.054908}{{\em Phys. Rev. C}
  {\bfseries 89} no.~5, (2014) 054908},
  \href{http://arxiv.org/abs/1402.7348}{{\ttfamily arXiv:1402.7348 [hep-ph]}}.

\end{thebibliography}\endgroup
\bibliographystyle{utphys}
\end{document}